\documentclass[6pt,twocolumn,showpacs,preprintnumbers,amsmath,amssymb,prb]{revtex4}

\usepackage{color}
\usepackage{graphicx}
\usepackage{dcolumn}
\usepackage{bm}

\renewcommand{\Im}{\mathfrak{Im}\,}
\renewcommand{\Re}{\mathfrak{Re}\,}
\renewcommand{\vec}[1]{\mathbf{#1}}
\newcommand{\mean}[1]{\left\langle #1 \right\rangle}

\begin{document}


\title{Nonequilibrium Green's functions approach to strongly correlated \\few-electron quantum dots}

\author{K. Balzer}
\affiliation{Institut f\"ur Theoretische Physik und Astrophysik, Christian-Albrechts-Universit\"at Kiel, Leibnizstrasse 15, 24098 Kiel, Germany}
\email{balzer@theo-physik.uni-kiel.de}
\author{R. van Leeuwen}
\affiliation{Department of Physics, University of Jyv\"askyl\"a, Survontie 9, 40014 Jyv\"askyl\"a, Finland}
\author{N.E. Dahlen}
\author{A. Stan}
\affiliation{Zernike Institute of Advanced Materials, University of Groningen, Nijenborgh 4, 9747 AG Groningen, The Netherlands}
\author{M. Bonitz}{
\affiliation{Institut f\"ur Theoretische Physik und Astrophysik, Christian-Albrechts-Universit\"at Kiel, Leibnizstrasse 15, 24098 Kiel, Germany}

\date{\today}

\begin{abstract}
The effect of electron-electron scattering on the equilibrium properties of few-electron quantum dots is investigated by means of nonequilibrium Green's functions theory. The ground and equilibrium state is self-consistently computed from the Matsubara (imaginary time) Green's function for the spatially inhomogeneous quantum dot system whose constituent charge carriers are treated as spin-polarized. To include correlations, the Dyson equation is solved, starting from a Hartree-Fock reference state, within a conserving (second order) self-energy approximation where direct and exchange contributions to the electron-electron interaction are included on the same footing.

We present results for the zero and finite temperature charge carrier density, the orbital-resolved distribution functions and the self-consistent total energies and spectral functions for isotropic, two-dimensional parabolic confinement as well as for the limit of large anisotropy---quasi-one-dimensional entrapment. For the considered quantum dots with $N=2$, $3$ and $6$ electrons, the analysis comprises the crossover from Fermi gas/liquid (at large carrier density) to Wigner molecule or crystal behavior (in the low-density limit).
\end{abstract}


\maketitle

\section{\label{sec:intro}Introduction}
In the recent decades finite quantum systems have become an intensively studied subject-matter. Particular interest is due to electrons in quantum dots \cite{jacak98} (QD) or wells, forming \textit{artificial atoms} \cite{ashoori96} with molecule-like behavior and novel spectral and dynamical properties. In contrast to \textit{real atoms}, such new properties arise from dimensionality reduction and natural scale-differences as QDs embedded within semiconductor heterostructures generate 
charge carrier motion on typically nanometer length scales. The collective, optical and transport properties of QDs are examined by experimental \cite{brocke03} and theoretical \cite{baer04,szafran04,indlekofer05} research activities in dependence on various dot parameters and geometries. For an overview see e.g.~Refs.~\cite{jacak98,banyai93,reimann02}. Many ground state calculations are available in the literature which are based on different methods---exact numerical diagonalization \cite{kvaal07,ciftja06,jauregui93}, self-consistent Hartree-Fock \cite{yannou07,reusch01,ludwig08}, configuration interaction \cite{rontani06} and quantum Monte Carlo \cite{egger99}. Extensions to finite temperatures and to QD properties in (transversal) magnetic fields are to be found in Refs.~\cite{szafran04,yannou07,filinov01}.

Typical charge densities in QD devices vary over a large range \cite{ciftja06}---from macroscopic charge carrier ensembles to mesoscopic, few- and even single-electron quantum dots. However, injecting only a small integral number of electrons into the dot reveals system properties that sensitively depend on the charge carrier number and are thus externally controllable, e.g.~by gate voltage or tip-electrode field variation or local mechanical strain (band mismatch). On the other hand, the quantum dot state is governed by the interplay of quantum and spin effects, the Coulomb repulsion between the carriers and the strength of the dot confinement. This generally leads to strong electron-electron correlation i.e.~collision or scattering effects the influence of which on the many-particle state is very important in the behavior at zero and finite temperature.

The two-dimensional (2D) $N$-electron quantum dot Hamiltonian to be considered is
\begin{eqnarray}\label{ham}
 \hat{H}_{e}&=&\sum_{i=1}^{N}\left(-\frac{\hbar^2}{2 m_e^*}\nabla_{\!i}^2+\frac{m_e^*}{2} \omega_0^2 \vec{r}_i^2\right)+\sum_{i<j}^{N}\frac{e^2}{4\pi\varepsilon\, r_{ij}}\;,\;\;\;
\end{eqnarray}
where the effective electron mass is denoted by $m_e^*$, the frequency $\omega_0$ adjusts the (isotropic) parabolic confinement strength, $e$ is the elementary charge and $\varepsilon$ is the background dielectric constant. The vectors $\vec{r}_i$ are the single charge carrier coordinates with respect to the quantum dot center and $r_{ij}=|\vec{r}_i-\vec{r}_j|$. The density in the QD is controllable by the confining potential which directly affects the relative electron-electron interaction strength and, tuned towards low carrier densities, continuously leads to formation of so-called electron (Wigner) molecules \cite{egger99,reusch01} or crystal-like behavior \cite{filinov01}. Melting processes owing to an increased temperature cause weakening and finally preventing of such solid-like structure formation. Also, with the restriction to Eq.~(\ref{ham}), the present analysis neglects nonideality effects such as defects and well-width fluctuations, see e.g.~Refs.~\cite{filinov04,bracker05}.

The objective of the present work is to analyze the electron correlations in the QD system (\ref{ham}) on the pathway from Fermi gas/liquid towards strongly correlated Wigner molecule behavior, i.e.~during the (density) delocalization-localization transition. To this end, in Sec.~\ref{sec:theory}, we present the finite temperature formalism of nonequilibrium Green's functions (NEGF) theory which is applied by a threefold motivation: (i)~the NEGFs allow for a consistent and conserving treatment of Coulomb correlations, (ii)~previous NEGF approaches to inhomogeneous QDs incorporate to our knowledge no strong carrier-carrier coupling, and (iii), in contrast to other methods, strong importance lies in the possibility for the direct extension of the approach to nonequilibrium situations with e.g.~time-dependent gate-voltage variations, quantum transport phenomena or optical switching\cite{banyai93}.

Using the NEGF technique, the properties of the investigated spin-polarized $N$-electron quantum dot in thermodynamic equilibrium follow from the self-consistently obtained imaginary time (Matsubara) Green's function. Such an approach has also more recently shown to give accurate results for real atoms and molecules, see Refs.~\cite{dahlen05,dahlen06,dahlen07}. The extension of the nonequilibrium Green's function ansatz from traditional applications on quasi-homogeneous quantum systems \cite{bonitz96,binder97,kwong98} to spatial inhomogeneity is thereby the major goal of the present analysis.

The theoretical part in Sec.~\ref{sec:theory} is followed by a detailed description of the iteration technique used to numerically solve the Dyson equation (in Hartree-Fock \textit{and} second Born approximation) to self-consistency. The results are discussed in Sec.~\ref{sec:results}. The starting point is the limit of large anisotropy where in Eq.~(\ref{ham}) we consider the limit $\omega_0^2{\vec r}_i^2\rightarrow\omega^2_{x,0} x^2_i+\omega_{y,0}^2 y^2_i$ with $\omega_{y,0}\gg\omega_{x,0}$, see Sec.~\ref{subsec:results1D}. In the case of $N=3$ (quantum dot lithium) and $6$ electrons, the charge carrier density, the orbital-resolved distribution functions and total energies are computed for different values of interaction strength and temperature. Also, we demonstrate that at finite temperatures, the second Born (correlation) corrections to the mean-field treatment yield significant density changes in an intermediate regime whereas in the high- and low temperature limit the electron density is only less affected by correlations. In Sec.~\ref{subsec:results2D}, we extend the calculations to isotropic 2D confinement and analogously report on ground state results for $N=2$ electrons (quantum dot helium) which are compared with exact and Hartree-Fock results of Ref.~\cite{reusch01}. Moreover, the computation of the charge carrier spectral function\cite{kadanoff62} $a(\omega)$ allows in Sec.~\ref{subsec:spectralfct} for a collision induced renormalization of the Hartree-Fock energy spectrum. This is of high relevance for the optical properties of the few-electron QD. Sec.~\ref{sec:conclude} gives a final discussion.

\section{\label{sec:theory}Theory}
For characterization and quantum mechanical treatment of the $N$-electron dot system (\ref{ham}) it is convenient to introduce the coupling (or Wigner) parameter $\lambda$ which relates the characteristic Coulomb energy $E_C=e^2/(4\pi\varepsilon l_0^*)$ to the confinement energy $E^*_0=\hbar\omega_0\;$:
\begin{eqnarray}\label{cp}
 \lambda=\frac{E_C}{E^*_0}=\frac{e^2}{4\pi\varepsilon\,l_0^* \hbar\omega_0}=\frac{l_0^*}{a_B}\;,
\end{eqnarray}
with $l_0^*=\sqrt{\hbar/(m^*_e\omega_0)}$ being the characteristic single-electron extension in the QD and $a_B$ the effective electron Bohr radius. Using the replacement rules $\{\vec{r}_i\rightarrow \vec{r}_i/l_0^*,\,E\rightarrow E/E_0^*\}$, Hamiltonian (\ref{ham}) transforms into the dimensionless form
\begin{eqnarray}\label{haml}
 \hat{H}_{\lambda}&=&\frac{1}{2}\sum_{i=1}^{N}(-\nabla_{\!i}^2+\vec{r}_i^2)\,+\,\sum_{i<j}^{N}\frac{\lambda}{r_{ij}}\;.
\end{eqnarray}
For coupling parameters $\lambda\ll1$, the quantum dot electrons will be found in a Fermi gas- or liquid-like state, whereas in the limit $\lambda\rightarrow\infty$, it is $l_0^*\gg a_B$, and quantum effects vanish in favor of classical interaction dominated charge carrier behavior \cite{ludwig08}. In the case of moderate coupling ($\lambda\gtrsim 1$) quantum dots with spatially well localized carrier density can be formed. Further, in addition to $N$ and $\lambda$, the system is characterized by the QD temperature $\beta^{-1}=k_B T$ which will be measured in units of the confinement energy $E_0^*$.

\subsection{\label{subsec:2ndquant}Second quantization representation}
Introducing carrier annihilation (creation) operators $\hat{\psi}^{(\dagger)}(\vec{r})$ with action at space point $\vec{r}$, the second-quantized form of Hamiltonian $\hat{H}_{\lambda}$, Eq. (\ref{haml}), is
\begin{eqnarray}\label{sq}
 \hat{H}_\lambda\!&=&\!\int\!\textup{d}^2\!r\,\hat{\psi}^\dagger(\vec{r})\,h^0(\vec{r})\,\hat{\psi}(\vec{r})\\
        &&+\,\frac{1}{2}\!\int\!\!\!\int\!\textup{d}^2\!r\,\textup{d}^2\! \bar{r}\,\hat{\psi}^\dagger(\vec{r})\,\hat{\psi}^\dagger(\bar{\vec{r}})\,\frac{\lambda}{\sqrt{(\vec{r}-\bar{\vec{r}})^2}}\,\hat{\psi}(\bar{\vec{r}})\,\hat{\psi}(\vec{r})\;,\nonumber
\end{eqnarray}
where $h^0(\vec{r})=(-\nabla^2+\vec{r}^2)/2$ denotes the single-particle energy and the second term in~(\ref{sq}) describes the electron-electron interactions. The field operators $\hat{\psi}^{(\dagger)}(\vec{r})$ satisfy the fermionic anti-commutation relations ${[\hat{\psi}(\vec{r}),\hat{\psi}^\dagger(\bar{\vec{r}})]}_{+}=\delta(\vec{r}-\bar{\vec{r}})$ and ${[\hat{\psi}^{(\dagger)}(\vec{r}),\hat{\psi}^{(\dagger)}(\bar{\vec{r}})]}_{+}=0$, where $[\hat{A},\hat{B}]_{+}=\hat{A} \hat{B}+\hat{B} \hat{A}$.

Ensemble averaging in Eq.~(\ref{sq}) directly gives rise to the one-particle nonequilibrium Green's function which is defined as
\begin{eqnarray}\label{1pngf}
 G(1,2)=-\frac{i}{\hbar}\mean{T_{\cal C}[\hat{\psi}(1)\hat{\psi}^\dagger(2)]}\;,
\end{eqnarray}
and is a generalization of the one-particle density matrix [which is recovered from $G$ in the limit of equal time arguments $t_1=t_2$, see e.g.~Ref.~\cite{kadanoff62}]. The used nomenclature is $1=(\vec{r}_1,t_1)$ and the expectation value (ensemble average) reads $\langle{\hat A}\rangle=\textup{Tr}\,{\hat{\rho}\hat A}$. The two times $t_{1}$ and $t_{2}$ entering $G(1,2)$ arise in the Heisenberg picture of the field operators and vary along the complex Schwinger/Keldysh contour ${\cal C}=\{t\in\mathbb{C}\,|\,\Re\,t\in[0,\infty]\,,\,\Im\,t\in[-\beta,0]\}$ where $T_{\cal C}$ denotes time-ordering on $\cal C$, see e.g. Refs.~\cite{dahlen06} and \cite{kadanoff62}. Note, that in the remainder of this work we use $\hbar=1$. The advantage of using the NEGF is that it allows for equal access to equilibrium and nonequilibrium averages at finite temperatures and that quantum many-body approximations can be systematically included by diagram expansions \cite{dahlen05}, see Secs.~\ref{subsec:negf} and \ref{sec:simu}. Moreover, most dynamic (spectral) and thermodynamic information \cite{kadanoff62} is contained in the NEGF, cf.~Sec.~\ref{sec:results}.

\subsection{\label{subsec:negf}Nonequilibrium Green's functions formalism}
The two-time nonequilibrium Green's function $G(1,2)$ obeys the Keldysh/Kadanoff-Baym equation (KBE) \cite{kadanoff62,bonitz96}
\begin{eqnarray}\label{kkbe}
 &&[i\,\partial_{t_1}-h^0(\vec{r}_1)]\,G(1,2)\nonumber\\
&=&\delta_{\cal C}(1-2)-\!\int_{\cal C} \textup{d}3\,W(1-3)\,G_{12}(13;23^+)\;,
\end{eqnarray}
and its adjoint equation [with interchanged time arguments \mbox{$t_1\leftrightarrow t_2$}]. On the right hand side of Eq.~(\ref{kkbe}) the (collision) integral runs over the full configuration space and the time domain spanned by the contour $\cal C$. Further, \mbox{$W(1-2)=\lambda\,\delta_{\cal C}(t_1-t_2)/\sqrt{(\vec{r}_1-\vec{r}_2)^2}$} is the instantaneous (time-local) electron-electron interaction and \mbox{$\delta_{\cal C}(1-2)=\delta_{\cal C}(t_1-t_2)\,\delta(\vec{r}_1-\vec{r}_2)$} with the time delta function being defined on the contour. $G_{12}(12;1'2')$ denotes the two-particle NEGF
\begin{eqnarray}\label{2pngf}
 G_{12}(12;1'2')=(-i)^2\mean{T_{\cal C}[\psi(1)\psi(2)\psi^\dagger({2'})\psi^\dagger({1'})]}\;,
\end{eqnarray}
where the short notation $3^+$ in Eq.~(\ref{kkbe}) indicates that the limit $t\rightarrow t_3+0$ is taken from above on the contour. In the integro-differential form (\ref{kkbe}), the KBE is not closed but constitutes the first equation of the Martin-Schwinger (MS) hierarchy \cite{martin59}. In order to decouple the hierarchy approximate expressions for the two-particle Green's function are introduced. E.g.~in a first order (spatially non-local) Hartree-Fock approach one substitutes $G_{12}(12;1'2')\rightarrow G(1,1')G(2,2')-G(1,2')G(2,1')$ which is known to preserve total energy and momentum but completely neglects correlations --- the former term leads over to the Hartree potential, the latter accounts for exchange. More generally, such \textit{conserving} approximations can be formulated in terms of a self-energy functional $\Sigma[G](1,2)$ which is defined by
\begin{eqnarray}\label{sigma}
 &&-i\int_{\cal C}\textup{d}3\,W(1-2)\,G_{12}(13;23^+)\nonumber\\
&=&\int_{\cal C}\textup{d}3\,\Sigma[G](1,3)\,G(3,2)\;.
\end{eqnarray}
Other, advanced conserving approximations, such as the second Born approximation (see Sec.~\ref{sec:simu}), can be systematically derived from a generating functional $\Phi[G]$ according to $\Sigma(1,2)=\delta \Phi[G]/\delta G(2,1)$, see e.g. Ref.~\cite{baym62}.} In addition to a specific MS hierarchy decoupling, the KBE (\ref{kkbe}) must be supplied with initial or boundary conditions. In this paper, we will use the Kubo-Martin-Schwinger conditions \cite{kubo5766,kadanoff62} $G(\vec{r}_1 t_1,2)|_{t_1=0}=-G(\vec{r}_1\,0-i \beta,2)$ and $G(1,\vec{r}_2 t_2)|_{t_2=0}=-G(1,\vec{r}_2\,0-i \beta)$.

In the case of thermodynamic equilibrium, where without loss of generality the electron system~(\ref{ham}) is time-independent for \mbox{$\Re t_{1,2}\leq0$}, $G(1,2)$ has no real-time dependence but extends on the imaginary contour branch $[-i\beta,0]$ only. We define the corresponding Matsubara (imaginary time) Green's function $G^M$ with respect to the transformations $t_1-t_2\rightarrow i\tau$ ($\tau\in[-\beta,\beta]$) and $G\rightarrow-iG^M$, i.e.
\begin{eqnarray}\label{mgf}
 G^M\!(\vec{r}_1,\vec{r}_2;\tau)=-i G(1,2)\;,
\end{eqnarray}
which only depends on the time difference $t_1-t_2$, $t_{1,2}\in[-i\beta,0]$ and is anti-periodic in the inverse temperature $\beta$, compare with definition (\ref{1pngf}). Using expressions (\ref{sigma}) and (\ref{mgf}) in the KBE (\ref{kkbe}) leads to the general form of the Dyson equation \cite{dahlen05} for the spin-polarized QD system (\ref{sq})
\begin{eqnarray}\label{deq}
 &&[-\partial_\tau-h^0(\vec{r}_1)]\,G^M\!(\vec{r}_1,\vec{r}_2;\tau)\\
&=&\delta(\tau)+\!\int\!\!\textup{d}^2\bar{r}\int_0^\beta\!\!\textup{d}\bar{\tau}\,\Sigma^M_\lambda(\vec{r}_1,\bar{\vec{r}};\tau-\bar{\tau})\,G^M\!(\bar{\vec{r}},\vec{r}_2;\bar{\tau})\;,\nonumber
\end{eqnarray}
with the anti-periodic Matusbara self-energy $\Sigma^M_\lambda(\vec{r}_1,\vec{r}_2;\tau)$. Note that the Dyson equation in this form is exact and that many-body approximations enter via $\Sigma^M_\lambda[G^M]$.

Eq.~(\ref{deq}) is the central equation which will be applied in the subsequent Secs.~\ref{sec:simu} and \ref{sec:results} to investigate the effect of carrier-carrier correlations in the $N$-electron quantum dot. However, as the self-energy $\Sigma^M_\lambda$ appears as a functional of the Matsubara Green's function $G^M\!(\vec{r}_1,\vec{r}_2;\tau)$, a self-consistent solution of the Dyson equation is required to accurately characterize the equilibrium QD state. The corresponding numerical technique is developed in the next section.

\section{\label{sec:simu}Simulation technique}
In this section, we discuss the computational scheme of solving the Dyson equation for the 2D few-electron quantum dot specified by Eq.~(\ref{haml}). Thereby, we proceed in two steps: First, we solve Eq.~(\ref{deq}) at the Hartree-Fock (HF) level, see Sec.~\ref{subsec:HF}, and, second, we incorporate correlations within the $\Phi$-derivable second order Born approximation, see Sec.~\ref{subsec:2ndB}. Throughout, we represent $G^M$ in the $\tau$-domain rather than solving the Dyson equation in frequency space where $G^M\!(\vec{r}_1,\vec{r}_2;\omega)=\int_0^\beta\!\textup{d}\tau\,G^M\!(\vec{r}_1,\vec{r}_2;\tau)\,e^{i\omega \tau}$ can be obtained by analytic continuation, see e.g.~Ref.~\cite{ku02}.

\subsection{\label{subsec:HF}Hartree-Fock at zero and finite temperatures}
At mean-field level, the solution of the Dyson equation (\ref{deq}) is fully equivalent to the Hartree-Fock self-consistent field method \cite{yannou07,echenique07} at finite temperatures $\beta^{-1}$. Hence, we primarily resort to standard HF techniques and will recover the uncorrelated Matsubara Green's function, denoted $G^0(\vec{r}_1,\vec{r}_2;\tau)$, at the end of this section.

The Hartree-Fock approach leads to an effective one-particle description of the QD and gives a first estimate of exchange effects. However, as an independent-electron approximation, it does not include correlations, i.e.~the HF total energy is given by $E_{\mathrm{HF}}^0=E_{\mathrm{exact}}-E_{\mathrm{corr}}$. With respect to the second quantized Hamiltonian of Eq.~(\ref{sq}), the effective HF Hamiltonian is obtained by approximately replacing the four field operator product entering the interaction term by sums over products $\hat{\psi}^\dagger\hat{\psi}$ weighted by the generalized carrier density matrix $\rho(\vec{r},\bar{\vec{r}})$. This is consistent with the mean-field approximation for the two-particle Green's function as given in Sec.~\ref{subsec:negf} and leads to
\begin{eqnarray}
 \hat{H}_\lambda=\!\int\!\!\!\int\!\textup{d}^2 r\,\textup{d}^2 \bar{r}\,\hat{\psi}^\dagger(\vec{r})[h^0(\vec{r})\delta(\vec{r}-\bar{\vec{r}})+\Sigma^0_\lambda(\vec{r},\bar{\vec{r}})]\hat{\psi}(\vec{r})\;,\nonumber\\
\label{HFham}
\end{eqnarray}
with the Hartree-Fock self-energy
\begin{eqnarray}\label{HFse}
 \Sigma^0_\lambda(\vec{r},\bar{\vec{r}})=\!\int\!\textup{d}^2 r'\,\frac{\lambda\rho(\vec{r}',\vec{r}')}{\sqrt{(\vec{r}'-\vec{r})^2}}\,\delta(\vec{r}-\bar{\vec{r}})-\frac{\lambda\rho(\vec{r},\bar{\vec{r}})}{\sqrt{(\vec{r}-\bar{\vec{r}})^2}}\;.\;\;
\end{eqnarray}
Here, the first (second) term constitutes the Hartree (Fock or exchange) contribution.

Computationally convenient is the introduction of a basis representation for the electron field operator according to
\begin{eqnarray}
\label{fobexp}
\hat{\psi}^{(\dagger)}(\vec{r})=\sum_{i}\varphi^{(*)}_i(\vec{r})\,\hat{a}^{(\dagger)}_{i}\;,\hspace{1pc}i\in\{0,1,2,\ldots\}\;, 
\end{eqnarray}
where the one-particle wave functions (orbitals) $\varphi_i(\vec{r})$ form an orthonormal complete set and $\hat{a}^{(\dagger)}_{i}$ denotes the annihilation (creation) operator of a particle on the level $i$. At this stage, the QD system (\ref{HFham}) can be transformed into the matrix representation $h_{\lambda,ij}=h^0_{ij}+\Sigma^0_{\lambda,ij}$ with the single particle quantum numbers $i$ and $j$, $h_{ij}$ being the electron HF total energy, $h^0_{ij}$ the single-particle (kinetic plus confinement) energy and $\Sigma^0_{\lambda,ij}$ the electron self-energy in mean-field approximation. More precisely, we have
\begin{eqnarray}
\label{HFhamm}
 h_{\lambda,ij}&=&h^0_{ij}+\Sigma^0_{\lambda,ij}\;,\\
\label{H0term}
 h_{ij}^0&=&\frac{1}{2}\int\!\textup{d}^2 r\,\varphi_i^*(\vec{r})(-\nabla^2+\vec{r}^2)\varphi_j(\vec{r})\;,\\
\label{HFterm}
 \Sigma^0_{\lambda,ij}&=&\lambda\sum_{kl} (w^{}_{ij,kl}-w^{}_{il,kj}) \rho_{kl}(\beta)\;,
\end{eqnarray}
with the finite (zero) temperature charge carrier density matrix $\rho_{ij}(\beta)=\langle\hat{a}^\dagger_i \hat{a}_j\rangle$ (in the limit $\beta\rightarrow\infty$) in the grand canonical ensemble and the two-electron integrals $w_{ij,kl}^{}$ defined as
\begin{eqnarray}\label{2ei}
w_{ij,kl}^{}&=&\!\int\!\!\!\int \textup{d}^2r\,\textup{d}^2\bar{r}\, \frac{\varphi^*_i(\vec{r})\,\varphi^*_k(\bar{\vec{r}})\,\varphi_j(\vec{r})\,\varphi_l(\bar{\vec{r}})}{\sqrt{(\vec{r}-\bar{\vec{r}})^2+\alpha^{2}}}\;.
\end{eqnarray} 
Using $\alpha\rightarrow0$, the integrals in $w_{ij,kl}^{}$ can be performed analytically in 2D but, in the limit of large anisotropic confinement (quasi-1D quantum dot), a truncation parameter $0<\alpha\ll1$ is needed to regularize the (bare) Coulomb potential at $|\vec{\vec{r}-\bar{\vec{r}}}|=0$ keeping $w_{ij,kl}^{}$ finite, see e.g.~Ref.~\cite{ciftja06}.
Alternatively, the parameter $\alpha$ adjusts a confining potential in the perpendicular dimension and allows (at small $r_{ij}$) for a transversal spread of the wave function \cite{jauregui93}. For the specific choice of the parameter $\alpha$, see Sec.~\ref{sec:results}.

Using standard techniques, we iteratively solve the self-consistent Roothaan-Hall equations \cite{roothaan51} for the Hartree-Fock Hamiltonian $h_{\lambda,ij}$, Eq.~(\ref{HFhamm}),
\begin{eqnarray}\label{rheq}
 \sum_{k=0}^{n_b-1}h_{\lambda,ik}\,c_{kj}-\epsilon^0_{j}\,c_{ij}=0\;,
\end{eqnarray}
which at finite dimension $n_b\times n_b$ ($i=0,1,\ldots,n_b-1$) yield the numerically exact eigenfunctions (HF orbitals) expanded in the form $\phi_{\lambda,i}(\vec{r})=\sum_{j=0}^{n_b-1} c_{ji}\,\varphi_j(\vec{r})$, $c_{ij}\in\mathbb{R}$, the corresponding energy spectrum (HF eigenvalues) $\epsilon_i^0$ and the chemical potential $\mu^0$. Consequently, the $N$-electron quantum dot system is fully characterized by the solution $\phi_{\lambda,i}(\vec{r})$, e.g.~its charge carrier density is given by
\begin{eqnarray}\label{H0density}
 \rho^0(\vec{r})&=&\sum_{i=0}^{n_b-1}f(\beta,\epsilon_i^0-\mu^0)\,\phi_{\lambda,i}(\vec{r})\\
&=&\sum_{i=0}^{n_b-1}f(\beta,\epsilon_i^0-\mu^0)\sum_{j=0}^{n_b-1}c_{ji}\,\varphi_j(\vec{r})\;, \nonumber
\end{eqnarray}
where $f(\beta,\epsilon_i^0-\mu^0)$ denotes the Fermi-Dirac distribution.

For numerical implementation of the mean-field problem (\ref{rheq}), we have chosen the Cartesian (2D) harmonic oscillator states
\begin{eqnarray} \label{ost}
\varphi_{m,n}(\vec{r})&=&\frac{e^{-(x^2+y^2)/2}}{\sqrt{2^{m+n}\,m!\,n!\,\pi}}\,\,{\cal H}_m(x)\,{\cal H}_n(y)\;,
\end{eqnarray}
with single-electron quantum numbers $i=(m,n)$, $\vec{r}=(x,y)$ in units of the oscillator length $l_0^*$, the Hermite polynomials ${\cal H}_m(x)$ and $(m+1)$-fold degenerate energies $\epsilon_{m,n}=m+n+1$, $m,n\in\{0,1,2,\ldots\}$. In the 1D quantum dot limit, these states reduce to the one-dimensional oscillator eigenfunctions $\varphi_{m}(x)=(2^m m! \sqrt{\pi})^{-1/2}\,e^{-x^2/2}\,{\cal H}_m(x)$.

As mentioned before, the self-consistent Hartree-Fock result generates an uncorrelated Matsubara Green's function $G^0(\vec{r}_1,\vec{r}_2;\tau)$  which yields the same observables. For instance, the $N$-electron density of Eq.~(\ref{H0density}) is recovered from $\rho^0(\vec{r})=\,G^0(\vec{r},\vec{r};0^-)$---the energy contributions are discussed separately in Sec.~\ref{sec:results}. Expanding $G^0$ in terms of the obtained HF basis $\phi_{\lambda,i}(\vec{r})$ according to
\begin{eqnarray}\label{gfexp}
 G^0(\vec{r}_1,\vec{r}_2;\tau)=\sum_{ij}\phi^*_{\lambda,i}(\vec{r}_1)\,\phi_{\lambda,j}(\vec{r}_2)\,g^0_{ij}(\tau)\;,
\end{eqnarray}
with associated $\tau$-dependent real matrix elements $g_{ij}^0(\tau)$, leads to the identity
\begin{eqnarray}\label{HFgf}
 g^0_{ij}(\tau)=\delta_{ij}\,f(\beta,\epsilon_i^0-\mu^0)\,e^{\tau(\epsilon_i^0-\mu^0)}\;,\;
\end{eqnarray}
which is (band) diagonal only in the HF orbital basis and solves the Dyson equation in mean-field approximation
\begin{eqnarray}\label{deqHF}
 [-\partial_\tau-\vec{h}^0-\vec{\Sigma}^0_\lambda]\,\vec{g}^0(\tau)=\delta(\tau)\;.\;
\end{eqnarray}
Here, the time-independent matrices $(\vec{h}^0)_{ij}=h^0_{ij}$ and $(\vec{\Sigma}^0_\lambda)_{ij}=\Sigma^0_{\lambda,ij}$ are defined in correspondence to Eqs.~(\ref{H0term}) and (\ref{HFterm}) with $\varphi_i$ being replaced by $\phi_{\lambda,i}$, the Hartree-Fock Green's function being denoted as $(\vec{g}^0(\tau))_{ij}=g^0_{ij}(\tau)$, and the charge carrier density matrix due to Eq.~(\ref{HFterm}) reads $\rho_{ij}(\beta)=g_{ij}^0(0^-)$ with notation $0^-$ denoting the limit from below on the contour $\cal C$. Further, it is $(\vec{a}\,\vec{b})_{ij}=\sum_{k}a_{ik}\,b_{kj}$. Note, that in Eq.~(\ref{2ei}) also the two-electron integrals are to be transformed into their HF representation, and that in the following bold-typed expressions as introduced in Eq.~(\ref{deqHF}) denote matrices with respect to the HF basis $\phi_{\lambda,i}(\vec{r})$.

\subsection{\label{subsec:2ndB}Solving the self-consistent Dyson equation beyond the Hartree-Fock level}
In this subsection, we focus on electron-electron correlation corrections to the self-consistent Hartree-Fock reference state \cite{dahlen05} determined by $G^0(\vec{r}_1,\vec{r}_2;\tau)$. The idea is to start from the Dyson equation (\ref{deq}) in HF orbital representation
\begin{eqnarray}\label{deqm}
 [-\partial_\tau-\vec{h}^0]\,\vec{g}^M\!(\tau)=\delta(\tau)+\!\int_0^\beta\!\!\!\textup{d}\bar{\tau}\,\vec{\Sigma}^M_\lambda\!(\tau-\bar{\tau})\,\vec{g}^M\!(\bar{\tau})\;,\;
\end{eqnarray}
with the full, time-dependent Matsubara self-energy $(\vec{\Sigma}^M_\lambda(\tau))_{ij}=\Sigma^M_{\lambda,ij}(\tau)$ and the equilibrium Green's function $(\vec{g}^M(\tau))_{ij}=g^M_{ij}(\tau)$, both obtained by applying the orbital expansion of Eq.~(\ref{gfexp}).
An explicit approximate expression for ${\vec{\Sigma}}^M_\lambda$ including correlation effects is introduced below, cf.~Eqs.~(\ref{msecorr}-\ref{2ndBse}).

First, we discuss the general solution scheme for Eq.~(\ref{deqm}). However, we will not consider it in this form. Instead, we integrate Eq.~(\ref{deqm}) inserting Eq.~(\ref{HFgf}) and applying the anti-periodicity property of $\vec{g}^M\!(\tau)$. This leads to the integral form of the Dyson equation
\begin{widetext}
\begin{eqnarray}\label{deqif}
 \vec{g}^M\!(\tau)-\int_0^\beta\!\!\!\textup{d}\bar{\bar{\tau}}\!\int_0^\beta\!\!\!\textup{d}\bar{\tau}\,\vec{g}^0(\tau-\bar{\bar{\tau}})\,\vec{\Sigma}^r_\lambda[\vec{g}^M](\bar{\bar{\tau}}-\bar{\tau})\,\vec{g}^M\!(\bar{\tau})&=&\vec{g}^0(\tau)\;,\;\\
\label{mse}
\vec{\Sigma}^r_\lambda[\vec{g}^M](\tau)&=&\vec{\Sigma}^M_\lambda[\vec{g}^M](\tau)-\delta(\tau)\,\vec{\Sigma}^s_\lambda\;,
\end{eqnarray}
\end{widetext}
where the expression $\vec{\Sigma}^r_\lambda(\tau)$ according to definition (\ref{mse}) implicates the total Matsubara self-energy reduced by the initial (steady-state) mean-field $\vec{\Sigma}^s_\lambda=\vec{\Sigma}^0_\lambda[\vec{g}^0(0^-)]$ which is not a functional of the full (correlated) Green's function $\vec{g}^M(\tau)$. In addition, the single-particle energy $\vec{h}^0$ has already been absorbed in the HF reference state $\vec{g}^0(\tau)$ and thus does not appear explicitly in Eq.~(\ref{deqif}). For a more detailed derivation of Eq.~(\ref{deqif}) see Appendix.

We highlight, that the integral form of the Dyson equation can be parameterized by the second index $j\in\{0,1,\ldots,n_b-1\}$ of the Matsubara Green's function $g^M_{ij}(\tau)$ since the matrix multiplications on the left hand side of Eq.~(\ref{deqif}) do not affect this index. Hence, at a fixed Matsubara self-energy and discretized $\tau$-interval $[-\beta,\beta]$, Eq.~(\ref{deqif}) allows for reinterpretation as a set of $n_b$ independent (but typically large-scale) linear systems of the form
\begin{eqnarray}\label{leqs}
{\cal A}\,{\cal X}^{(j)}&=&{\cal B}^{(j)}\;,
\end{eqnarray}
where the unknown quantity and the inhomogeneity are $({\cal X}^{(j)})_{ip}=g_{ij}^M(\tau_p)$ and $({\cal B}^{(j)})_{ip}=g_{ij}^0(\tau_p)$, respectively. The coefficient matrix $({\cal A})_{ip,jq}={\alpha}_{ij}(\tau_p,\tau_q)$ is defined by the expression (convolution integral)
\begin{eqnarray}\label{convint}
{\alpha}_{ij}(\tau,\bar{\tau})&=&\delta_{ij}\delta(\tau-\bar{\tau})\\
&&-\sum_{k=0}^{n_b-1}\!\int_{0}^\beta\!\!\!\textup{d}\bar{\bar{\tau}}\,{g}^0_{ik}(\tau-\bar{\bar{\tau}})\,{\Sigma}^r_{\lambda,kj}(\bar{\bar{\tau}}-\bar{\tau})\;,\nonumber
\end{eqnarray}
in which the integral over $\bar{\tau}$ in the Dyson equation (\ref{deqm}) vanishes due to its replacement by the matrix multiplication ${\cal A}\,{\cal X}^{(j)}$. In more detail, we need to specify the time-discretization of the Matsubara Green's function undertaken in Eq.~(\ref{leqs}): First, due to the anti-periodicity property of $G^M$, we can restrict ourselves to solve Eq.~(\ref{deqm}) on the negative $\tau$-interval $[-\beta,0]$. This specific choice originates from the fact that in the limit $\tau\rightarrow0^-$ the density matrix is obtained from $\vec{g}^M(\tau)$. Second, the numerical treatment must take into account the time-dependence of $G^M(\tau)$. From Eq.~(\ref{HFgf}) it follows that the Green's function is essentially peaked around $\tau=0$ and $\pm\beta$. Thus, not an equidistant grid but a uniform power mesh\cite{upmesh} (UPM) is adequate to represent the Green's function---this method is also used in Refs.~\cite{ku02,dahlen05}. With a total number of $n_m$ mesh points the dimensionality of the linear system ${\cal A}\,{\cal X}^{(j)}={\cal B}^{(j)}$ becomes $n_b n_m\times n_b n_m$. As stated above, Eq.~(\ref{leqs}) can only be processed for a fixed self-energy $\vec{\Sigma}^r_\lambda[\vec{g}^M](\tau)$. This means, that in order to provide a self-consistent solution of the Dyson equation, we have to iterate the procedure by computing, at each step, a new self-energy  from the current $\vec{g}^M(\tau)$. This loop is then repeated until convergence.

So far, we have not specified a certain self-energy approximation. In Eq.~(\ref{mse}), one generally can split $\vec{\Sigma}^M_\lambda(\tau)$ into a mean-field and a correlation part, i.e.
\begin{eqnarray}\label{msecorr}
 \vec{\Sigma}_{\lambda}^M[g^M](\tau)=\delta(\tau)\,\vec{\Sigma}_{\lambda}^0[g^M(0^-)]+\vec{\Sigma}^{\mathrm{corr}}_{\lambda}[g^M](\tau)\;,\;\;\;
\end{eqnarray}
where the Hartree-Fock contribution,
\begin{eqnarray}
\label{HFtermdeqm}
 \Sigma^0_{\lambda,ij}&=&\lambda\sum_{kl} (w^{}_{ij,kl}-w^{}_{il,kj}) g_{kl}^M(0^-)\;,
\end{eqnarray}
is exact (compare with Eq.~(\ref{HFterm})) and the correlation part $\vec{\Sigma}^{\mathrm{corr}}(\tau)$, at the second Born level, is given by
\begin{eqnarray}\label{2ndBse}
 \Sigma^{\mathrm{corr}}_{\lambda,ij}(\tau)&=&\!\!-\sum_{klmnrs}w_{ik,ms}^{}(w^{}_{lj,rn}-w^{}_{nj,rl})\\
 &&\hspace{3pc}\times\,g^M_{kl}(\tau)\,g^M_{mn}(\tau)\,g^M_{rs}(-\tau)\;.\;\;\nonumber
\end{eqnarray}
Here, the first term denotes the direct contribution whereas the second one includes the exchange---for details see e.g. Refs.~\cite{kadanoff62,dahlen05}. Note, that the two-electron integrals $w_{ij,kl}^{}$ are given in their Hartree-Fock basis representation. Since the interaction potential $W(1,2)$ enters Eq.~(\ref{2ndBse}) in second order the present description of charge carrier correlations goes beyond the first Born approximation of conventional scattering theory.

When the self-consistency cycle reaches convergence, the matrix $\vec{g}^M(\tau)$ becomes independent of the initial state $\vec{g}^0(\tau)$ and, in configuration space, the correlated Matsubara Green's function of the QD system (\ref{haml}) follows as
\begin{eqnarray}
\label{mgfexpression}
 G^M(\vec{r}_1,\vec{r}_2;\tau)=\sum_{ij}\phi^*_{\lambda,i}(\vec{r}_1)\,\phi_{\lambda,j}(\vec{r}_2)\,g^M_{ij}(\tau)\;,
\end{eqnarray}
where the HF orbitals $\phi_{\lambda,j}(\vec{r})$ are those obtained in Sec.~\ref{subsec:HF}. Consequently, correlations are included via the $\tau$-dependent matrix elements $g_{ij}^M(\tau)$, $\tau\in[-\beta,0]$, which give access to the electron density via $\rho(\vec{r})=G^M(\vec{r},\vec{r};0^-)$. We note, that $G^0$ as obtained from the self-consistent HF calculation is only one possible reference (initial) state which can be used in the Dyson equation (\ref{deqif}). Also different types of uncorrelated Green's functions e.g. obtained from density-functional theory (DFT) or orbitals in local density approximation (LDA) are applicable if they satisfy the correct boundary conditions. For a recent discussion on the relevance for atoms and molecules see Ref.~\cite{dahlen05}.

In summary, the presented procedure is valid for arbitrary temperatures $\beta^{-1}$ and arbitrary coupling parameters $\lambda$. Thereby, the scope of numerical complexity is determined by the parameters $n_b$ (matrix dimension associated with the HF basis size) and $n_m$ (time-discretization on the UPM) which must be chosen with respect to convergence of the QD observables. Corresponding to Eq.~(\ref{deqif}) it has been found that particularly the particle number $N=\sum_{i=0}^{n_b-1}{g}_{ii}^M(0^-)$ and the correlation energy $E_{\mathrm{corr}}$ sensitively depend on $n_m$, cf. Eq.~(\ref{egycorr}) in Sec.~\ref{sec:results}.

\begin{table}
\begin{ruledtabular}
\begin{tabular}{cccccccc}
&  &  &  &  &  & & \\
$N\!=\!3$& (1D) &  &  &  &  & & \\
\hline
$\lambda$  & $E_{\mathrm{HF}}^0$ & $\mu^0$ & $E_{\mathrm{2ndB}}$ & $E_0$  & $E_{\mathrm{HF}}$ & $E_{\mathrm{corr}}$ & $E_{\mathrm{QMC}}$\\
\hline
$\beta\!=\!1$&  &  &  &  &  & & \\
\hline
$1$&  $8.173$ & $4.621$ & $8.201$ & $6.115$ & $2.321$ & $-0.235$ & $7.661$ \\
\underline{$2$}&  \underline{$10.066$} & $6.124$ & \underline{$10.215$} & $6.556$ & $4.405$ & $-0.747$ & $9.510$ \\
\hline
$\beta\!=\!2$&  &  &  &  &  & & \\
\hline
$1$&  $7.043$ & $4.670$ & $7.065$ & $5.027$ & $2.153$ & $-0.115$ & $6.761$ \\
\underline{$2$}&  \underline{$8.790$} & $6.169$ & \underline{$8.941$} & $5.311$ & $3.936$ & $-0.306$ & $8.603$ \\
$4$&  $11.732$ & $8.852$ & $11.918$ & $5.920$ & $6.303$ & $-0.304$ & $11.721$ \\
$6$&  $14.387$ & $11.231$ & $14.374$ & $6.712$ & $7.822$ & $-0.160$ & $14.403$ \\

$8$&  $16.790$ & $13.362$ & $16.747$ & $7.514$ & $9.354$ & $-0.120$ & $16.809$ \\
$10$&  $19.005$ & $15.294$ & $18.962$ & $8.257$ & $10.800$ & $-0.095$ & $19.034$ \\
\hline
$\beta\!=\!10$& (GS)  &  &  &  &  & & $E_{\mathrm{QMC}}^{\beta=5}$\\
\hline
$1$&  $6.615$ & $4.673$ & $6.591$ & $4.645$ & $1.987$ & $-0.042$ & $6.529$ \\
\underline{$2$}& \underline{$8.480$} & $6.173$ & \underline{$8.421$} & $4.966$ & $3.560$ & $-0.105$ & $8.371$ \\
$4$&  $11.667$ & $8.853$ & $11.578$ & $5.817$ & $5.917$ & $-0.156$ & $11.484$ \\
$6$&  $14.374$ & $11.243$ & $14.292$ & $6.710$ & $7.720$ & $-0.137$ & $14.161$ \\
$8$&  $16.787$ & $13.376$ & $16.721$ & $7.534$ & $9.296$ & $-0.110$ & $16.570$ \\
$10$&  $19.004$ & $15.298$ & $18.950$ & $8.285$ & $10.752$ & $-0.087$ & $18.791$ \\
\hline
 &  &  &  &  &  & & \\
$N\!=\!6$& (1D) &  &  &  &  & & \\
\hline
$\lambda$  & $E_{\mathrm{HF}}^0$ & $\mu^0$ & $E_{\mathrm{2ndB}}$ & $E_0$  & $E_{\mathrm{HF}}$ & $E_{\mathrm{corr}}$ & $E_{\mathrm{QMC}}$\\
\hline
$\beta\!=\!10$& (GS)  &  &  &  &  & & \\
\hline
$1$&  $27.600$ & $9.263$ & $27.519$ & $18.613$ & $9.028$ & $-0.123$ & --- \\
\underline{$2$}& \underline{$36.145$} & $12.195$ & \underline{$35.919$} & $19.976$ & $16.289$ & $-0.346$ & --- \\
$4$&  $50.960$ & $16.110$ & $50.440$ & $23.666$ & $27.384$ & $-0.609$ & --- \\
\hline
&  &  &  &  &  & & \\
\end{tabular}
\end{ruledtabular}
\caption{Different energy contributions in dependence on the coupling parameter $\lambda$ for the 'ground state' (GS, $\beta=10$) and equilibrium states ($\beta=2$ and $1$) of $N=3$ and $6$ spin-polarized electrons in a quasi-1D quantum dot. $\mu^0$ gives the chemical potential as obtained from the Hartree-Fock calculation with total energy $E_{\mathrm{HF}}^0$, see Sec.~\ref{subsec:HF}. $E_{\mathrm{2ndB}}$, $E_0, E_{\mathrm{HF}}$, and $E_{\mathrm{corr}}$ are computed from the correlated Green's function $\vec{g}^M(\tau)$. All energies are in units of $E_0^*$ and the underlined values pertain to the results shown in Figs.~\ref{Fig:2}-\ref{Fig:4} and Fig.~\ref{Fig:5}. For comparison, $E_{\mathrm{QMC}}$ denote the total energy obtained from quantum Monte Carlo (QMC) simulations, see also Fig.~\ref{Fig:QMC}.}
\label{table:1DN3}
\end{table}

\begin{figure}[t]
 \includegraphics[width=0.475\textwidth]{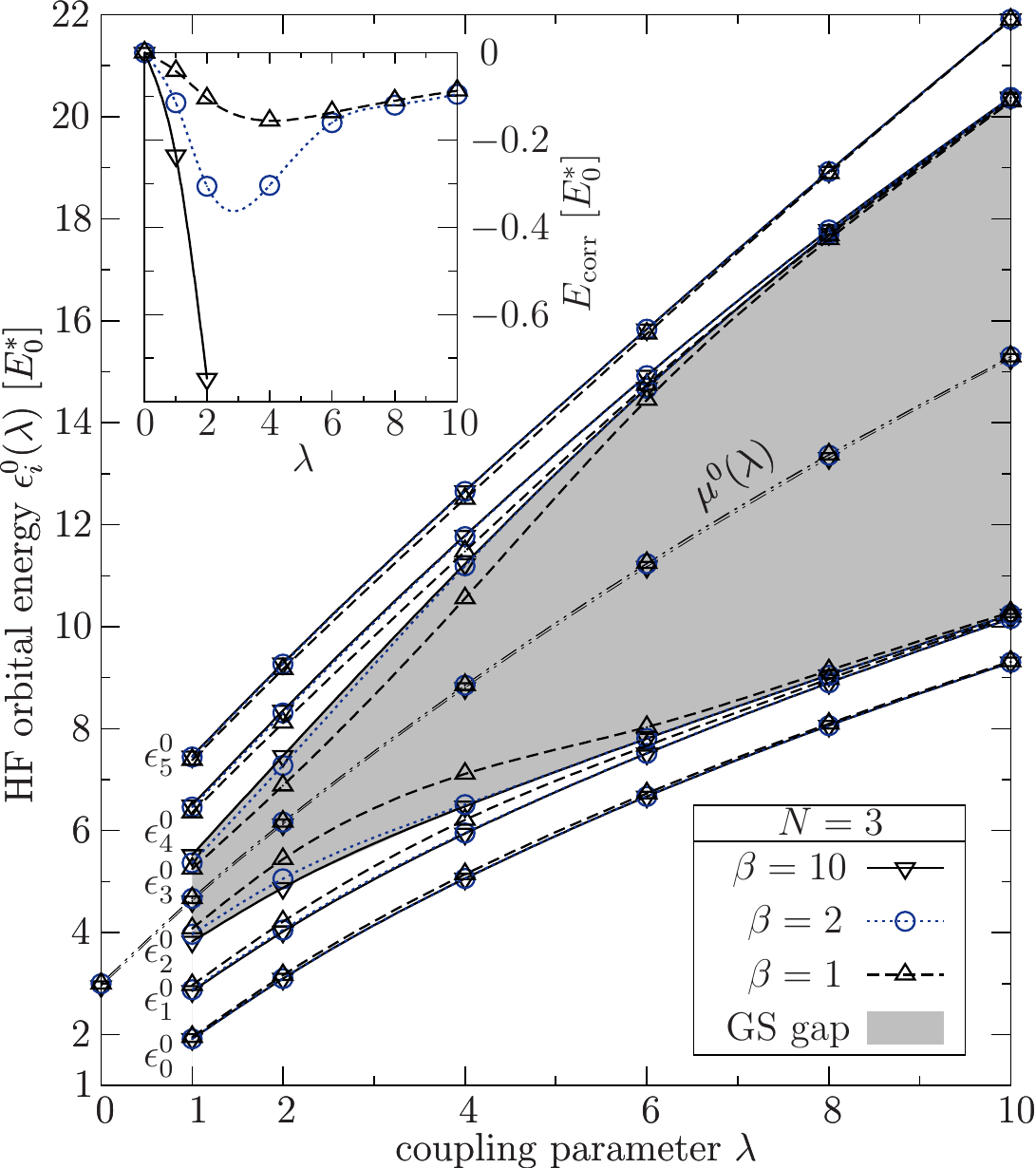}
\caption{(color online) The six energetically lowest HF orbital energies $\epsilon_i^0$ in dependence on $\lambda$ for the three-electron QD at different temperatures $\beta^{-1}$. The gray area indicates the ground state HOMO-LUMU gap between the occupied and unoccupied states. The chemical potential $\mu^0(\lambda)$ (double-dotted-dashed curves) is situated within this gray area. Inset: $\lambda$-depencence of the correlation energy $E_\mathrm{corr}$, see Table ~\ref{table:1DN3}.}\label{Fig:eval}
\end{figure}

\begin{figure}[t]
 \includegraphics[width=0.475\textwidth]{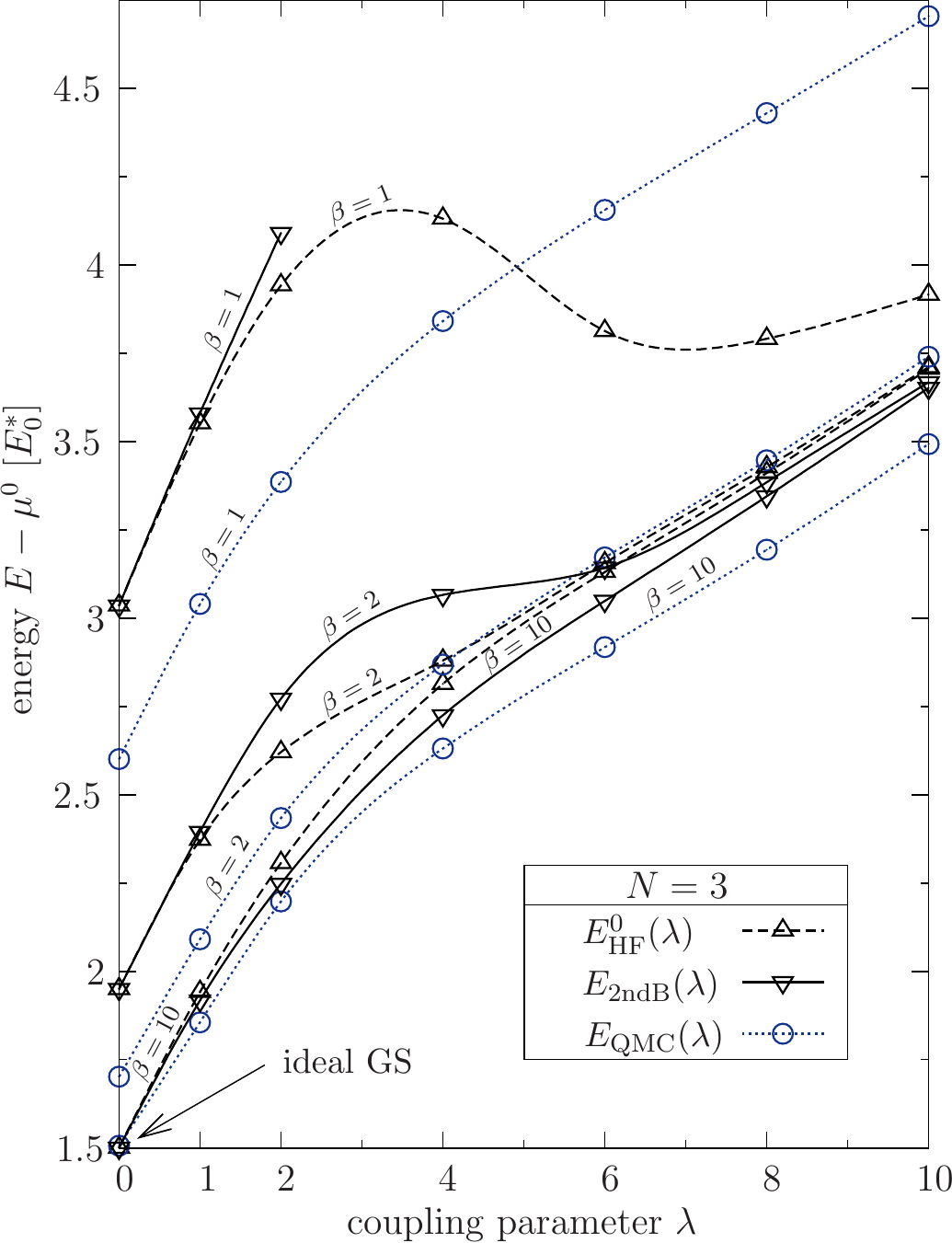}
\caption{(color online) Total energies in dependence on $\lambda$ and $\beta$ as given in Table \ref{table:1DN3} for the three-electron QD. Comparison of the [grand canonical] Green's function result (at HF and second Born level) with quantum Monte Carlo [canonical]. For $\lambda\equiv0$, the total energy can be analytically obtained from the (grand) canonical partition function of the noninteracting system according to standard formulas\cite{thesis07,tran01}.}\label{Fig:QMC}
\end{figure}

\section{\label{sec:results}Numerical Results}
In this section, we report on the numerical results for the few-electron quantum dots with $N=2$, $3$ and $6$ charge carriers. At that, we mainly focus on the energies and the (accumulated) single-carrier density and compare the influence of HF and second Born type self-energies, i.e.~Eq.~(\ref{HFtermdeqm}) versus Eqs.~(\ref{HFtermdeqm}) plus (\ref{2ndBse}).

The energies that contribute to the total energy of the QD system are, in addition to the single-particle (kinetic [$\vec{t}^0$] and confinement [$\vec{v}^0$]) energy $E_0=\textup{Tr}\,\vec{h}^0\,\vec{g}^M(0^-)=\textup{Tr}\,(\vec{t}^0+\vec{v}^0)\,\vec{g}^M(0^-)$, the mean-field Hartree-Fock and the correlation energy \cite{dahlen06} defined as
\begin{eqnarray}\label{egyHFHF}
E_{\mathrm{HF}}=\frac{1}{2}\textup{Tr}\,\vec{\Sigma}_\lambda^0\,\vec{g}^M(0^-)\;,
\end{eqnarray}
\begin{eqnarray}\label{egycorr}
E_{\mathrm{corr}}=\frac{1}{2}\int_{0}^\beta\!\!\textup{d}\tau\,\textup{Tr}\,\vec{\Sigma}_\lambda^{\mathrm{corr}}(-\tau)\,\vec{g}^M(\tau)\;.
\end{eqnarray}
The total energy is then $E_{\mathrm{2ndB}}=E_0+E_{\mathrm{HF}}+E_{\mathrm{corr}}$. For comparison, the total energy with respect to the HF Green's function $G^0(\vec{r}_1,\vec{r}_2;\tau)$ will be denoted by $E_{\mathrm{HF}}^0$. For evaluation of the two-electron integrals needed in Eqs.~(\ref{egyHFHF}) and (\ref{egycorr}), we have chosen the truncation parameter $\alpha=0.1$ in 1D and $\alpha\equiv0$ in 2D [no divergence of the integrals $w_{ij,kl}$, see Eq.~(\ref{2ei}) in Sec.~\ref{subsec:HF}].

Moreover, the HF orbital-resolved energy distribution functions (level occupation probabilities) $n_i(N;\lambda,\beta)$ are analyzed with respect to correlation induced scattering processes of particles into different energy levels. In general, it is
\begin{eqnarray}
 n_i=n_i(N;\lambda,\beta)=g^M_{ii}(0^-)\;,
\end{eqnarray}
which, in the case of vanishing correlations (\mbox{$G^M\rightarrow G^0$}), is just the Fermi-Dirac distribution, i.e.~$n_i=g^0_{ii}(0^-)=f(\beta,\epsilon_i^0-\mu^0)=f_i^0$, cf.~Eq.~(\ref{HFgf}).

\subsection{\label{subsec:results1D}Limit of large anisotropy (1D)}
When in Hamiltonian (\ref{ham}) the isotropic confinement of frequency $\omega_0$ is replaced by an anisotropic entrapment according to $\omega_{y,0}\gg\omega_{x,0}$, the QD charge carriers move effectively in one dimension. With the finite regularization parameter $\alpha=0.1$ we thereby allow for a small transversal extension (perpendicular to the $x$-axis). That is why we will call the system in this regime quasi-one-dimensional (quasi-1D).

In the following, let us first consider the 1D version of quantum dot lithium\cite{mikhailov02} ($N=3$) and, hereafter, a QD with $N=6$ confined electrons. For the respective HF and second Born calculations we throughout have used $n_b=30$ oscillator functions, see Sec.~\ref{subsec:HF}, and the number of mesh points $n_m(u,p)$ for the $\tau$-interval $[-\beta,0]$, discretized in Sec.~\ref{subsec:2ndB}, was varied between $60$ to $140$ in order to achieve convergence and preservation of particle number in the Dyson equation (\ref{deqif}). Table~\ref{table:1DN3} gives an overview of the relevant energies obtained at different coupling parameters and temperatures. Also, we included reference data from quantum Monte Carlo (QMC) simulations\cite{QMC} for the three-electron QD.

\begin{figure}[t]
\includegraphics[width=0.485\textwidth]{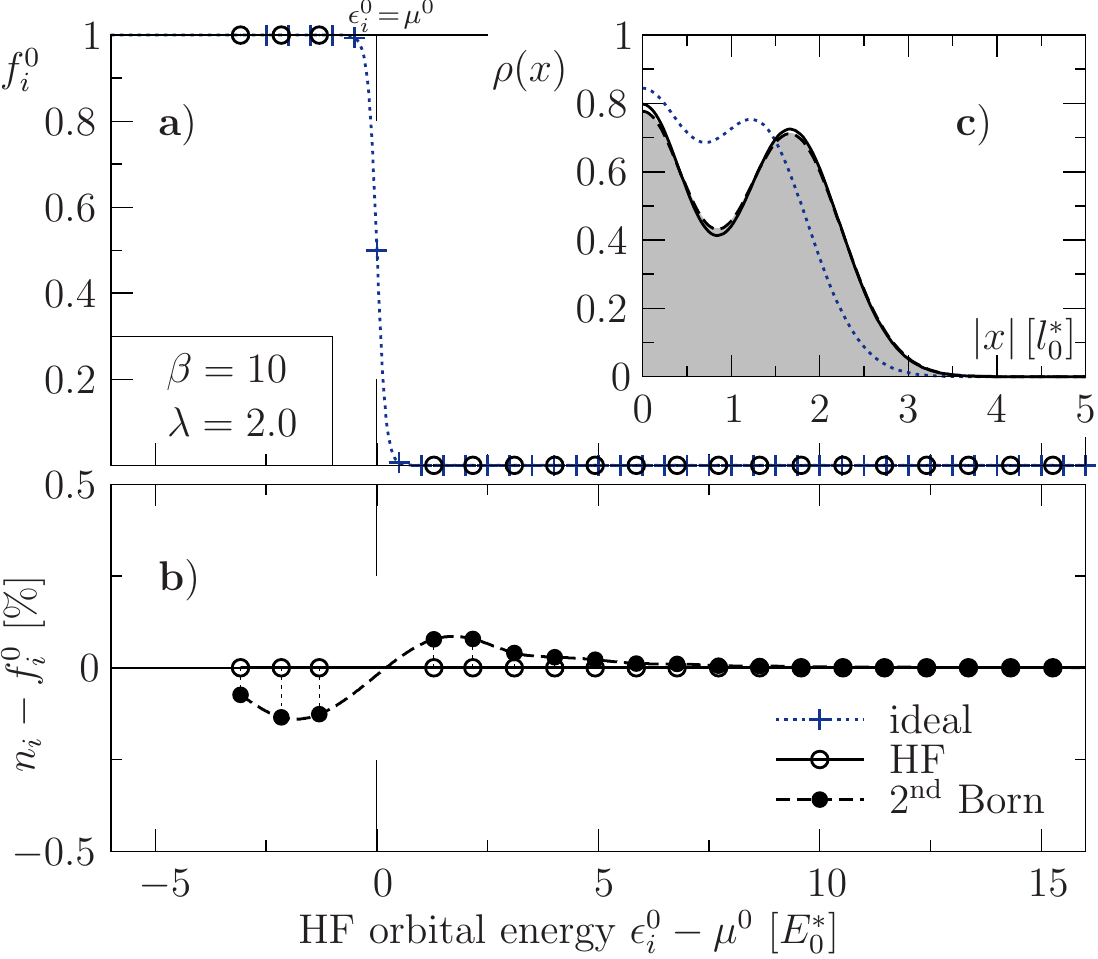}
\caption{(color online) Thermodynamic properties of $N=3$ electrons in the quasi-1D QD at $\beta=10$ and $\lambda=2.0$. \textbf{a}) Ideal and HF energy distribution functions $f_i^0$, \textbf{b}) change $n_i-f^0_i$ of the HF distribution due to correlations (in percent), and \textbf{c}) charge density profiles $\rho(x)$. The ideal (dotted line) and the HF result (solid line) are displayed together with the second Born approximation (dashed line).}\label{Fig:2}
\end{figure}
\begin{figure}[t]
\includegraphics[width=0.485\textwidth]{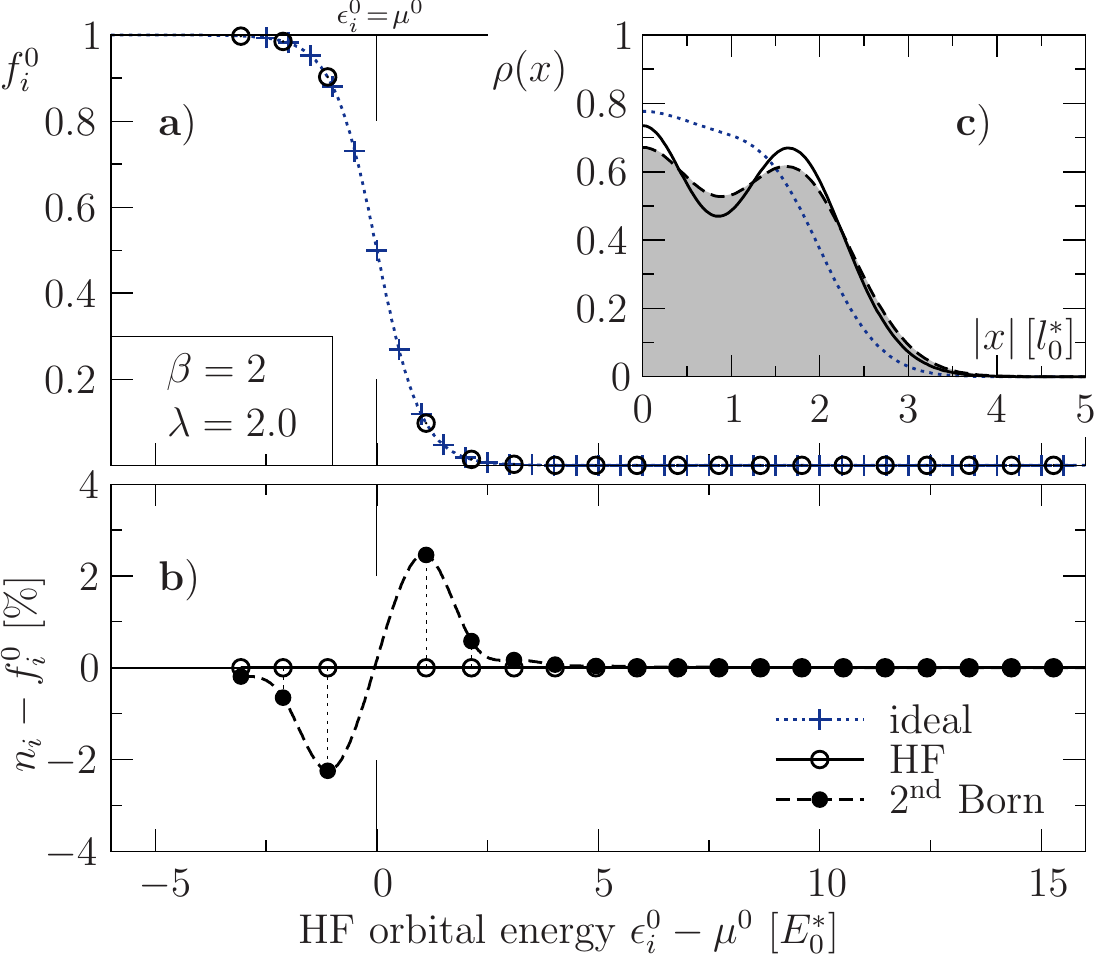}
\caption{(color online) Same as Fig.~\ref{Fig:2}, but for temperature $\beta=2$.}\label{Fig:3}
\end{figure}
\begin{figure}[t]
\includegraphics[width=0.485\textwidth]{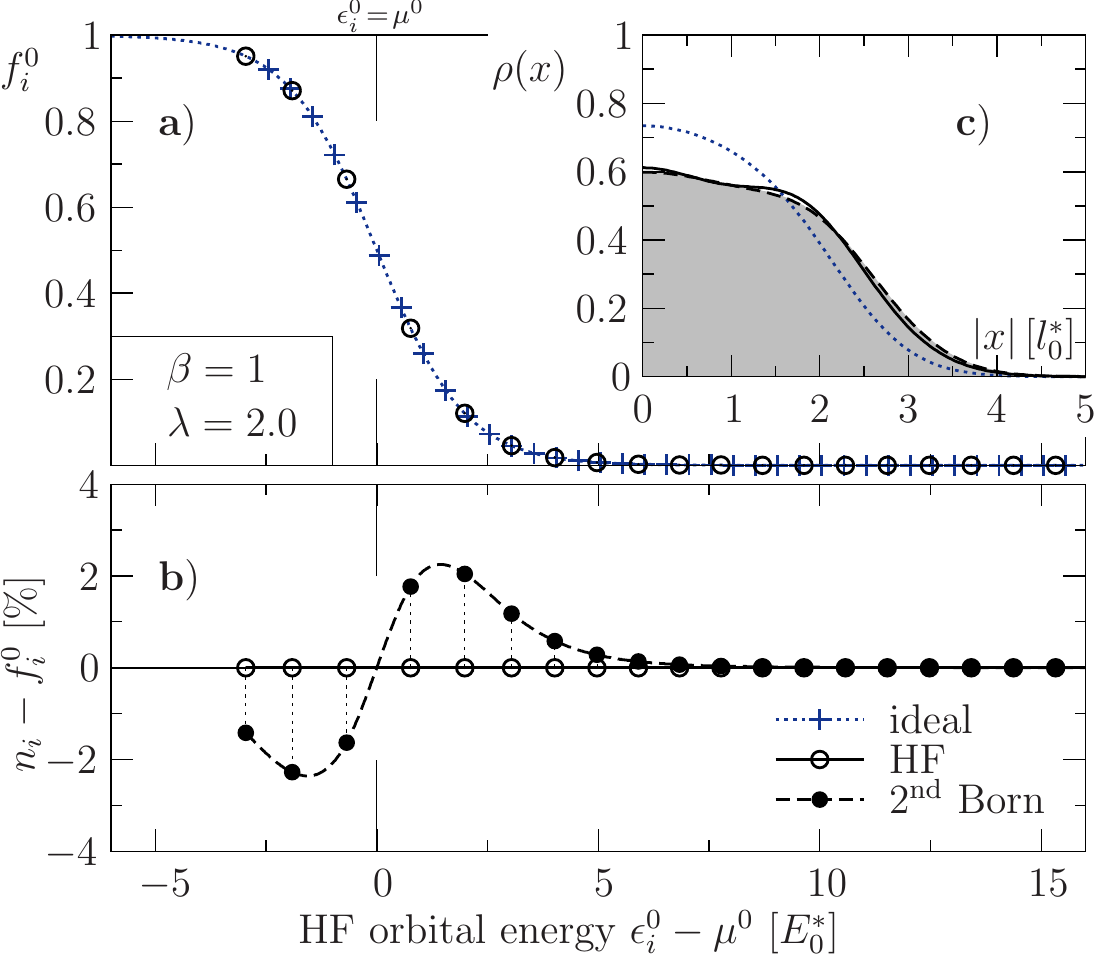}
\caption{(color online) Same as Fig.~\ref{Fig:2}, but for temperature $\beta=1$.}\label{Fig:4}
\end{figure}
\begin{figure}[t]
\includegraphics[width=0.485\textwidth]{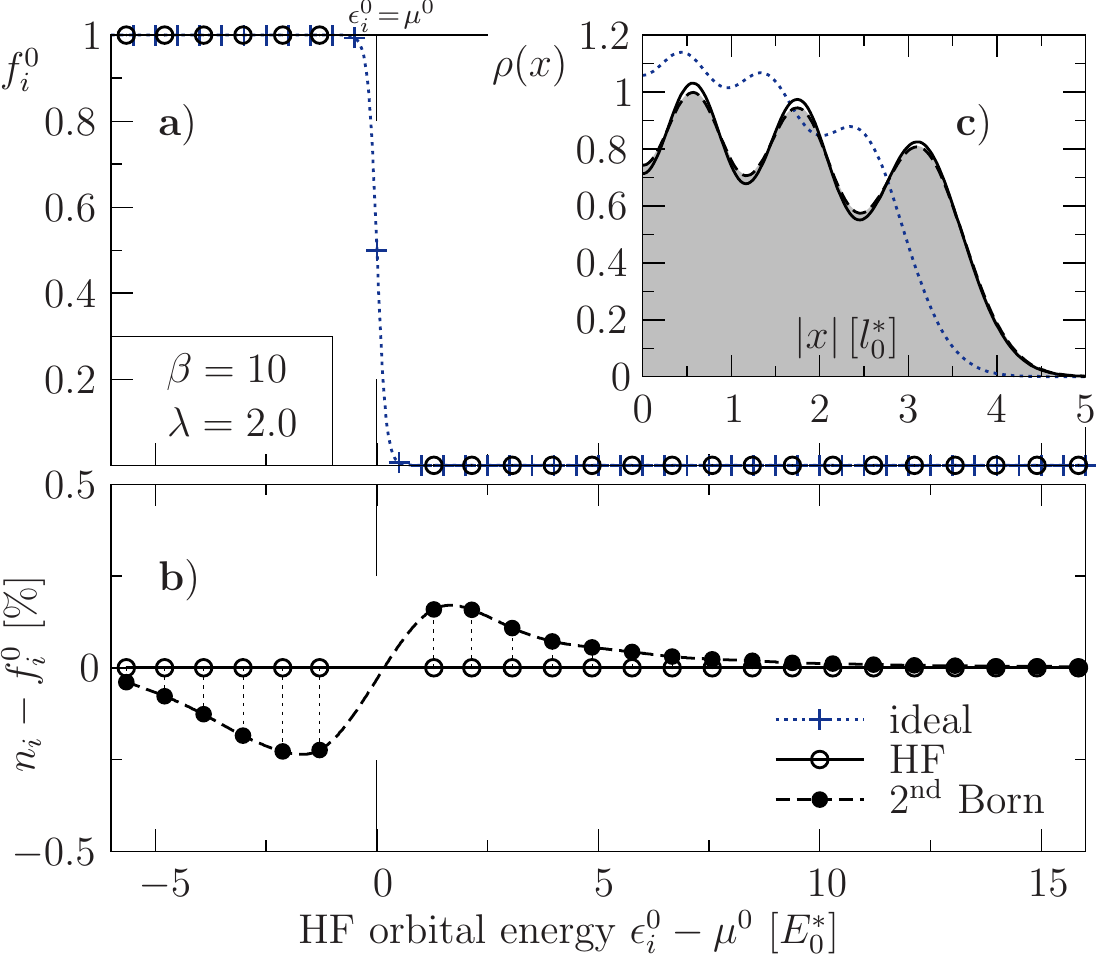}
\caption{(color online) Thermodynamic properties of $N=6$ charge carriers confined in the quasi-1D QD at $\beta=10$ and $\lambda=2.0$. \textbf{a}) Ideal and HF energy distribution functions $f_i^0$, \textbf{b}) change $n_i-f^0_i$ of the HF distribution due to correlations (in percent), and \textbf{c}) charge density profiles $\rho(x)$. Labeling as in Fig.~\ref{Fig:2}-\ref{Fig:4}.}\label{Fig:5}
\end{figure}

The low-energetic discrete orbital energies $\epsilon_i^0$ contributing to the HF reference state $\vec{g}^0(\tau)$ are shown in Fig.~\ref{Fig:eval} in dependence on $\lambda$ and $\beta$. For the quasi ground state (GS), $\beta=10$, the occupied states $i<3$ are energetically well separated from the unoccupied states $i\geq3$ by the HOMO-LUMO gap---the energy gap between the \underline{h}ighest \underline{o}ccupied (\underline{m}olecular or Hartree-Fock) \underline{o}rbital and the \underline{l}owest \underline{u}noccupied (\underline{m}olecular) \underline{o}rbital), see the gray area. For temperatures $\beta<10$, this gap is reduced eminently for moderate coupling around $\lambda\approx3$ (see the dotted and dashed lines), while, for $\lambda\rightarrow\infty$, the curves converge due to the strength of carrier-carrier interactions exceeding the incluence of thermal fluctuations. Moreover, the HF chemical potential $\mu^0(\lambda)$, situated within the HOMO-LUMO gap, is only slightly affected by $\beta$, compare the values in Table~\ref{table:1DN3}.

If we now include correlation effects, the HF spectra $\{\epsilon_i^0\}$ become renormalized---for the discussion see Sec.~\ref{subsec:spectralfct}. On the QD total energy, the influence of correlations is as follows: The correlation contribution $E_\mathrm{corr}$ is negative and increases with temperature but it is non-monotonic with regard to the coupling parameter $\lambda$, see Table~\ref{table:1DN3} and inset in Fig.~\ref{Fig:eval}. More precisely, the correlation effects are dominant at moderate coupling, between $\lambda=2$ and $6$, leading to lower total energies $E_{\mathrm{2ndB}}$ (compared to $E^0_\mathrm{HF}$) at temperature $\beta=10$ and to an energy increase for $\beta=2$ and $1$.

In general, both, the HF and second Born total energies well approach the corresponding [exact] QMC data independent of $\lambda$ and $\beta$, see Table~\ref{table:1DN3}. In order to properly compare the different approximations, the total energy is shown in Fig.~\ref{Fig:QMC} relatively to the mean-field chemical potential $\mu^0$. In the whole considered $\lambda$-regime, the quasi ground state energy $E_{\mathrm{HF}}^0$ (dashed curve for $\beta=10$) is downshifted by second Born corrections (solid curve) towards the energies obtained by QMC (dotted curve). At $\lambda=2$, we identify the best agreement of the correlated result with $E_\mathrm{QMC}$. In particular, for $\lambda\rightarrow0$, the three different total energies converge to the value $E-\mu^0=\frac{3}{2}$ of the ideal QD. At larger temperatures $\beta=2$ and $1$, the correlated Green's function $\vec{g}^M(\tau)$ leads, at moderate coupling around $\lambda\approx3$, to total energies that are significantly larger than the corresponding HF energies. This is consistent with the larger absolute values of the correlation energy $E_\mathrm{corr}$ given in Table~\ref{table:1DN3}. For stronger coupling $\lambda\gtrsim6$ (at $\beta=2$), $E_{\mathrm{2ndB}}$ then crosses the HF value in order to converge to the respective GS curve. The comparison with QMC is difficult at finite temperatures: We point out that (already at $\lambda=0$) there is a general shift due to the usage of different ensemble averages. Whereas QMC uses a canonical approach, the Green's function results emerge from a grand canonical picture. Nevertheless, close to the GS, Fig.~\ref{Fig:QMC} reveals a quite similar behavior in dependence on $\lambda$.

From Figs.~\ref{Fig:2}-\ref{Fig:4}~\textbf{a}), one gathers how the QD mean-field $\vec{\Sigma}^0_\lambda$ renormalizes the ideal equidistant energy spectrum $\epsilon_i=i+\frac{1}{2}$ of the noninteracting system ($\lambda\equiv0$), see the shifted HF energies $\epsilon_i^0$ (open circles) which exactly follow a Fermi-Dirac distribution according to Eq.~(\ref{HFgf}). Correlations due to $\vec{\Sigma}^\mathrm{corr}_\lambda$ now modify this statistics as can be seen from the quantity $n_i-f_i^0$ in Figs.~\ref{Fig:2}-\ref{Fig:4}~\textbf{b}), which measures the HF orbital-resolved deviation from the Fermi-Dirac distribution (in percent) and shows that charge carriers around $\mu^0$ are being scattered into higher HF orbitals (black circles). At larger temperatures, see e.g.~Fig.~\ref{Fig:3}~\textbf{b}), the change in the occupation probability $n_i$ exceeds $2$~\% for $\lambda=2.0$. Moreover, Pauli blocking inhibits energetically low lying electrons to essentially take part in the scattering process---consequently, $n_i-f_i^0$  is small for $\epsilon_i^0\ll\mu^0$.

\begin{table*}
\begin{ruledtabular}
\begin{tabular}{cccccccccccc}
$N=2$ & (2D) &&&&&&&&&\\
\hline
$\lambda$ & $E_{\mathrm{exact}}$  & $E_{\mathrm{HF}}^0$ & $\mu^0$ &  $E_{\mathrm{2ndB}}$ & $E_0$  & $E_{\mathrm{HF}}$ & $E_{\mathrm{corr}}$ & $\Delta_{\mathrm{HF}}^0$ [\%] & $\Delta_{\mathrm{2ndB}}$ [\%] & $\xi$ [\%] & $E_{\mathrm{QMC}}^{\beta=5}$\\
\hline
$1$   & ---      & $3.604$  & $2.885$       & $3.597$  & $3.023$      & $0.584$      & $-0.010$      & ---      & ---      & ---     & $3.591$ \\
$2$   & $4.142$  & $4.168$  & $3.393$       & $4.147$  & $3.082$      & $1.092$      & $-0.026$      & $0.638$  & $0.121$  & $80.8$  & $4.148$ \\
$4$   & $5.119$  & $5.189$  & $4.304$       & $5.117$  & $3.285$      & $1.901$      & $-0.069$      & $1.367$  & $0.039$  & $102.9$     & $5.123$ \\
\end{tabular}
\end{ruledtabular}
\caption{Different energy contributions in dependence on coupling parameter $\lambda$ for the ground state ($\beta=50$) of $N=2$ spin-polarized electrons in an isotropic 2D quantum dot. The exact energies for $\lambda=2$ and $4$ are quoted from Refs.~\cite{reusch01,merkt91} and arise by an exact diagonalization method. $E_\mathrm{QMC}$ gives quantum Monte Carlo reference data computed at the temperature $\beta=5$. $\Delta_x=|E_x-E_{\mathrm{exact}}|/E_{\mathrm{exact}}$ gives the relative error in \%. $\xi=(E_{\mathrm{HF}}^0-E_{\mathrm{2ndB}})/(E_{\mathrm{HF}}^0-E_{\mathrm{exact}})$ measures the correlation induced improvement of the total energy. All values are given in units of $E^*_0=\hbar\omega_0$. In particular, all three decimal places of the HF energies $E_{\mathrm{HF}}^0$ agree with Ref.~\cite{reusch01}.}
\label{table:2DN2}
\end{table*}

Figs.~\ref{Fig:2}-\ref{Fig:4}~\textbf{c}) visualize the inhomogeneous one-particle density $\rho(x)$ in the three-electron quantum dot ($\lambda=2.0$) in HF and second Born approximation. For comparison, we included also the density of the respective ideal system (blue dotted lines). Being symmetric around $x=0$, the density at low temperatures (Fig.~\ref{Fig:2} and \ref{Fig:3}~\textbf{c})) is three-fold modulated, and due to the electron-electron interactions the modulation in $\rho(x)$ is more intense compared to the ideal QD, where the oscillations originate from the Pauli principle only. Notably at $\beta=2$, correlations substantially weaken the density modulation and hence are important. At temperature $\beta=1$, the correlation effects are still present (see the quantity $n_i-f_i^0$) and reveal a more smooth, almost monotonic decay of the density with $x\rightarrow\infty$.

For the quasi 'ground state' ($\beta=10$) of the QD with six electrons, see Table~\ref{table:1DN3} and Fig.~\ref{Fig:5}, respectively, we observe similar properties as for the example of quantum dot lithium. Here, $\rho(x)$ becomes six-fold modulated and the quantum dot state can be interpreted as a \textit{Wigner chain} of six aligned charge carriers held together by the parabolic confinement. However, we note that no orbitals at $\lambda=2.0$ are degenerate, and that there is strong overlap of the single-carrier (HF) wave functions $\phi_{\lambda,i}(x)$. In contrast to the $N=3$ quantum dot, the influence of correlations on the equilibrium state $G^M$ (see again $n_i-f_i^0$) is stronger leading to considerable lowering of the total energy. This is consistent with an increased number of carriers and hence increased electron-electron collision probabilities, compare Figs.~\ref{Fig:2} and \ref{Fig:5}~\textbf{b}).

\subsection{\label{subsec:results2D} Isotropic quantum dot (2D)}
With no extra restriction on the carrier motion, Hamiltonian (\ref{ham}) describes a 2D QD with \textit{isotropic} parabolic confinement. Here, we report on the obtained Matsubara Green's function results for the special case of $N=2$ electrons, i.e.~for spin-polarized quantum dot helium. Thereby, we restrict ourselves to very low temperatures in order to compare with ground state data available in the literature\cite{reusch01}.

For the HF and second Born calculations, the results of which are shown in Table~\ref{table:2DN2}, the inverse temperature was set to $\beta=50$, and we used up to $n_b=40$ of the energetically lowest Cartesian oscillator functions, see Eq.~(\ref{ost}) in Sec.~\ref{subsec:HF}. Further, the uniform power mesh [$n_m(u,p)$] was chosen as in Sec.~\ref{subsec:results1D}, including essentially more than 100 grid points.

First, the unrestricted HF energies $E_\mathrm{HF}^0$ in Table~\ref{table:2DN2} exactly agree---to more than three decimal places---with the total energies computed in analogous manner by B.~Reusch et al., Ref.~\cite{reusch01}. This is a clear indication for the HF basis to be large enough. Second, applying the second Born approximation we are able to (essentially) improve these ground state results. The values of $E_\mathrm{2ndB}$, thereby, come quite close to the exact energies, which are obtained by numerical diagonalization\cite{merkt91}, and the data are also in good agreement with the QMC results. 

For instance, for coupling parameter $\lambda=2.0$, the inclusion of correlations reduces the relative error $\Delta_x$ by a factor of five---for definition of $\Delta_\mathrm{HF}$ and $\Delta_\mathrm{2ndB}$ see caption of Table~\ref{table:2DN2}. Hence, with respect to the HF solution, this means an improvement of the ground state total energy by about $\xi=(E_{\mathrm{HF}}^0-E_{\mathrm{2ndB}})/(E_{\mathrm{HF}}^0-E_{\mathrm{exact}})\approx80$~\%. For the strongly correlated case $\lambda=4.0$, the calculation slightly over-estimates the influence of correlations and leads to a total energy lower than the exact value. This, most probably, can be improved by increasing the number of mesh points for the $\tau$-interval which is crucial for the convergence of the correlation energy, see integral in Eq.~(\ref{egycorr}). Nevertheless, the exact ground state energy is considerably well approximated.

As a final remark, we note that our procedure does not restore the rotational symmetry of the Hamiltonian (\ref{ham}) into the two-dimensional solution. This is due the fact that we start from a symmetry broken \cite{yannou07} (HF) initial state $G^0$ when solving the Dyson equation to self-consistency. For the problem of restoring the symmetry, see e.g.~Ref.~\cite{giovannini07}. However, the presented results directly apply to QDs where impurities naturally avoid the rotational symmetry and hence lead to symmetry broken electron states.

\begin{figure}[b]
 \includegraphics[width=0.525\textwidth]{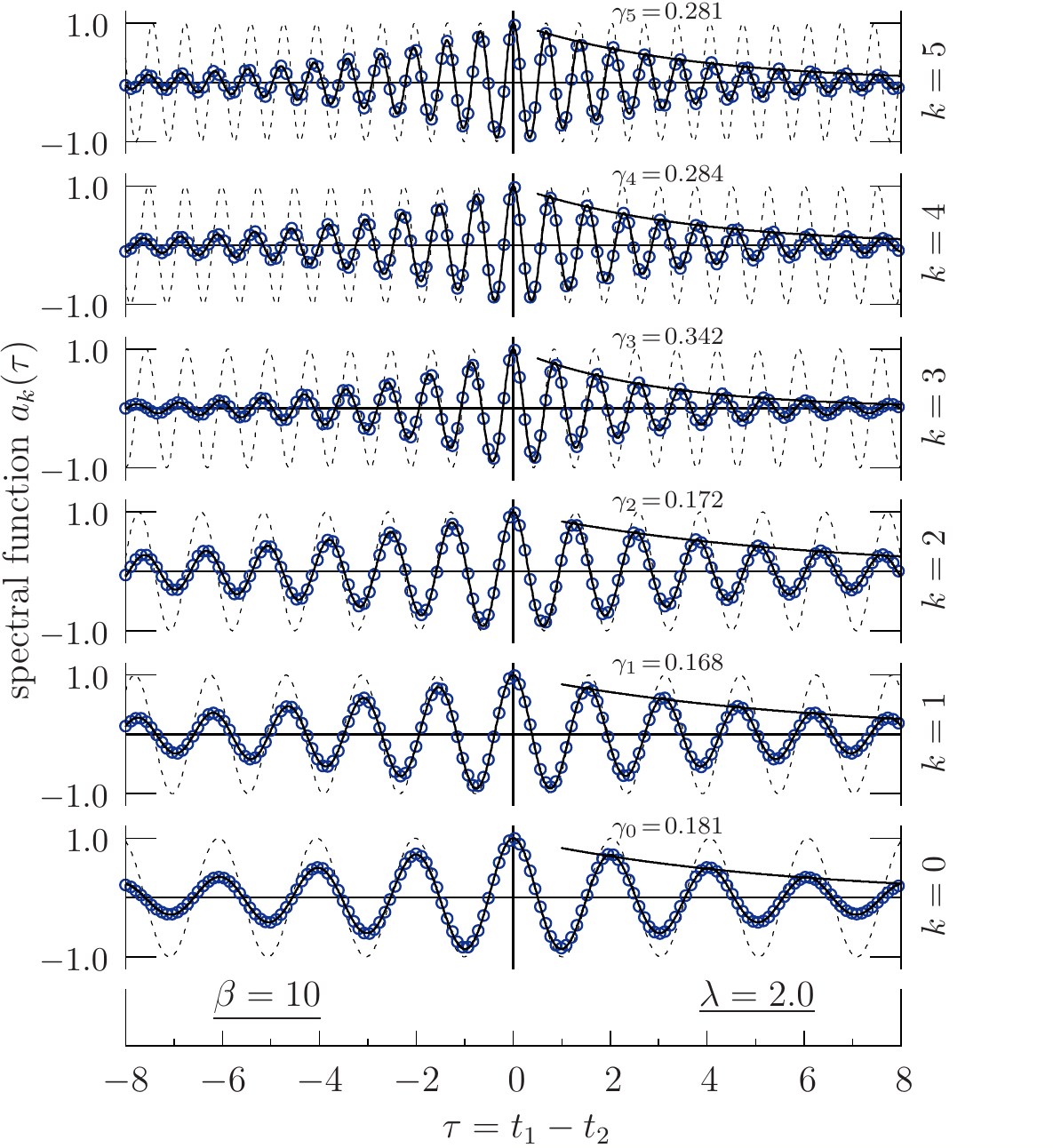}
\caption{(color online) Energetically lowest intraband spectral functions $a_k(\tau)$ (circles) for the one-dimensional quantum dot with $N=3$ charge carriers at $\beta=10$ and $\lambda=2.0$. The solid lines are the fitting results of the IHC model, Eq.~(\ref{spectralfct_IHC}), with exponential damping constants $\gamma_k=\eta_k\nu_k$ (solid exponential curves). The undamped oscillations (thin dashed curves in the background) correspond to the Hartree-Fock approximation, where no collisional broadening and thus no damping of $a_k(\tau)$ occurs.}\label{Fig.intrabandsf}
\end{figure}

\begin{figure*}[t]
 \includegraphics[width=0.975\textwidth]{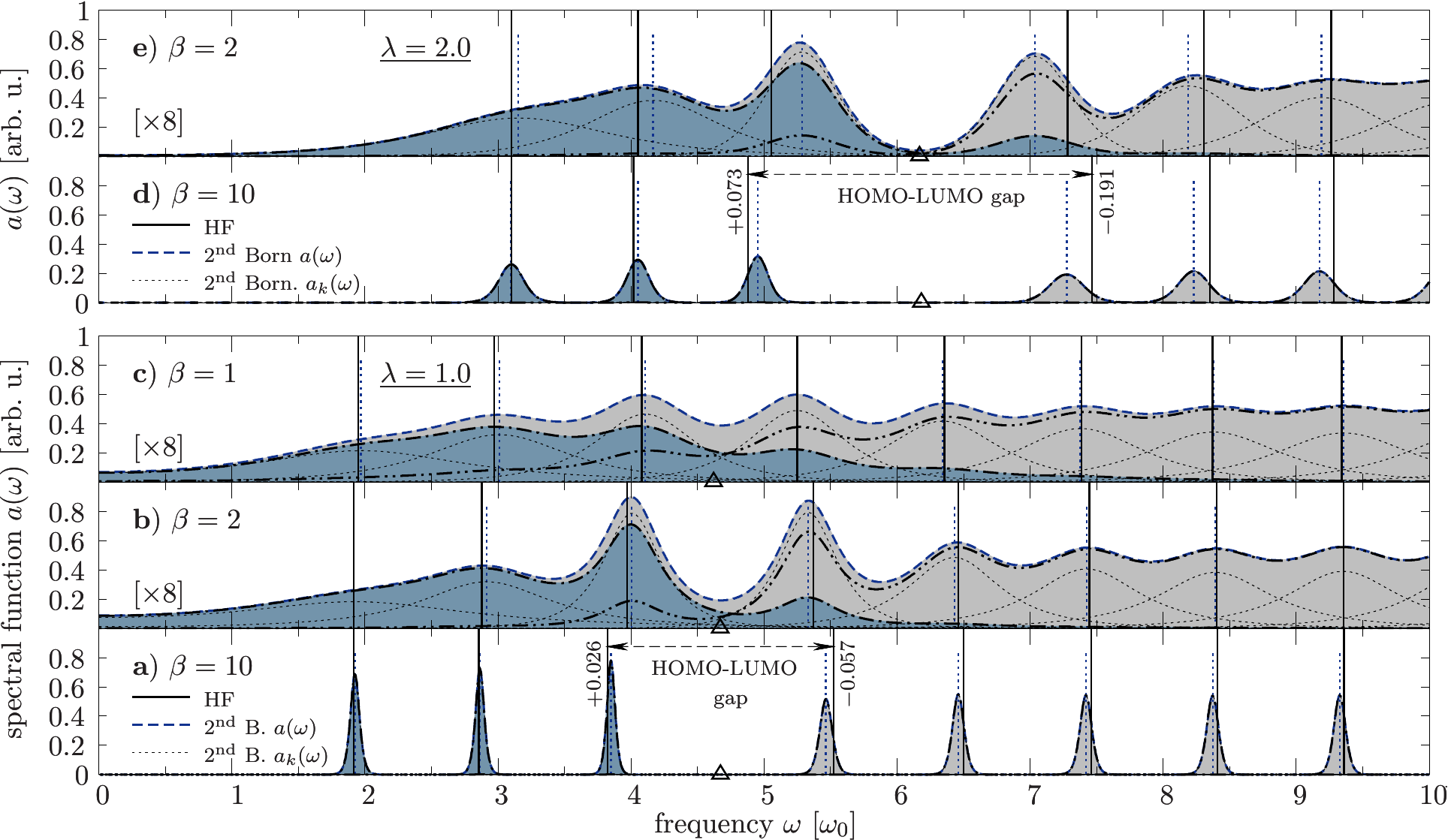}
\caption{(color online) Spectral function $a(\omega)$ (dashed curves filled gray) accumulated from the orbital-resolved functions $a_k(\omega)$ (thin dotted curves) for the quasi one-dimensional QD with $N=3$ electrons at different temperatures $\beta^{-1}$ as indicated---\textbf{a}) to \textbf{c}) $\lambda=1.0$, \textbf{d}) and \textbf{e}) $\lambda=2.0$. Note that the $y$-axis in \textbf{b}), \textbf{c}) and \textbf{e}) is stretched by factor $8$. The vertical solid lines denote the Hartree-Fock energies $\epsilon_k^0$, the dotted lines correspond to the maxima of the inverse hyperbolic cosines $a_k(\omega)$, Eq.~(\ref{spectralfct_IHC}). The numbers at the peak profiles in \textbf{a}) and \textbf{d}) denote the shift of the maxima in $a_{k=2,3}(\omega) $ with respect to the HF energies. The triangles on the abscissas give the position of the chemical potential $\mu^0$. Moreover, the dotted-dashed (double-dotted-dashed) curve show the (inverse) spectral weight---for more detailed information see text.}\label{Fig.sf}
\end{figure*}

\subsection{\label{subsec:spectralfct} Spectral function}
In HF approximation, the single-particle energy spectrum of the QD consists of discrete levels, see e.g.~$\{\epsilon_i^0\}$ for the three-carrier system considered in Fig.~\ref{Fig:eval}. When correlations are included the spectra generally turn into continuous functions of energy due to electron-electron scattering and provide additional informations such as finite line widths or temperature broadening. However, from the Matsubara Green's function, Eq.~(\ref{mgfexpression}), it is intricate to extract the correlated single-particle energy spectrum as its computation besides Fourier transformation usually involves Pad\'{e} approximations\cite{ku02}. The direct time-propagation of the equilibrium state $G^M(\vec{r}_1,\vec{r}_2;\tau)$ solving the Keldysh/Kadanoff-Baym equations (\ref{kkbe}) for the two-time NEGF
$G(1,2)=\theta(t_1-t_2)G^>(1,2)-\theta(t_2-t_1)G^<(1,2)$ initiates a more systematic approach. Here, the time-dependency of $G(1,2)$, which now extends also along the real part of the contour $\cal C$, provides access to the electron energies $\omega$ via the spectral function \cite{kadanoff62} $a(\omega)=\int_{-\infty}^{+\infty} \textup{d}\tau e^{i\omega\tau} a(\tau)$, where here $\tau$ is the difference of the two real time arguments in $G$.

The orbital-resolved carrier spectral function $\vec{a}(\tau)$ is given by
\begin{eqnarray}
 \vec{a}(\tau)=i \{\vec{g}^>(T-\tau,T+\tau)-\vec{g}^<(T-\tau,T+\tau)\}\;,
\end{eqnarray}
where $T\geq0$ is a specific point on the diagonal of the two-time plane ${\cal P}=[0,\infty]\times[0,\infty]$, $\tau=t_1-t_2\in[-\sqrt{2}T,\sqrt{2}T]$ denotes the relative time, and $\vec{g}^\gtrless(t_1,t_2)$ are the contour-ordered correlation functions with respect to the HF basis. Further, we identify the diagonal (offdiagonal) elements of matrix $\vec{a}(\tau)$ with the intraband (interband) spectral functions.

Well documented computational specifications for solving Eq.~(\ref{kkbe}) on ${\cal P}$, with initial condition $\vec{g}^M(\tau)$, are provided e.g.~by Refs.~\cite{dahlen06,dahlen07,thesis07}. We will not give a detailed description here. Nevertheless, it is instructive to note that, as the one-particle energy $\vec{h}^0$ includes no time-dependency, the dynamics of $G(1,2)$ can be obtained by propagation on the $t_{1,2}$ axes only [instead on whole ${\cal P}$] or by using the retarded Green's function\cite{banyai93} $G^R(1,2)=\frac{i}{\hbar}\langle[\hat{\psi}(1),\hat{\psi}^\dagger(2)]_{+}\rangle\theta(t_1-t_2)\,$.

As a result it turns out that the QD spectral function is not of Breit-Wigner type, i.e.~does not obey a distribution function $a_{kk}(\omega)=a_k(\omega)\propto\frac{1}{(\omega-\omega_k)^2+\gamma^2}$ as it follows from a quasi-particle (local approximation) ansatz\cite{bonitz99} with $a_k(\tau)=e^{i \epsilon_k\!\tau} e^{-\gamma \tau}$, single-particle energy $\epsilon_k$ and phenomenological damping $\gamma$. In contrast, the spectral function shows clear non-Lorentzian behavior, cf.~the circles in Fig.~\ref{Fig.intrabandsf} which shows the energetically lowest intraband spectral functions of the three-electron quantum dot discussed in Sec.~\ref{subsec:results1D}. To this end, at large $T\gg1$, we have adapted the computed spectral function $a_{k}(\tau)$ to an inverse hyperbolic cosine model \cite{haug98} (IHC)
\begin{eqnarray}
\label{spectralfct_IHC}
 a_{k}(\tau)=e^{i \omega_k\!\tau}\frac{1}{\cosh^{\eta_k}(\nu_k \tau)}\;,
\end{eqnarray}
which has been demonstrated to yield good results for Coulomb quantum kinetics, see Ref.~\cite{bonitz99}. The ansatz (\ref{spectralfct_IHC}) leaves open a set of three parameters $\{\omega_{k},\eta_{k},\nu_{k}\}$ (obtained by fitting), and, in accordance to the numerical data, (i)~ensures zero slope of $\Re a_{k}(\tau)$ at $\tau=0$ and (ii)~exhibits an exponential decay for large $\tau$ with a damping constant $\gamma_k=\eta_k \nu_k$. The former feature is especially missing in the quasi-particle picture. The solid curves in Fig.~\ref{Fig.intrabandsf} exemplify the good quality of the IHC model including properties (i)~and (ii)~and justify its usage.

In energy space, the $\gamma_k$-induced collisional broadening of the peaks in $a_k(\omega)$ can be shown to be again described by an inverse hyperbolic cosine\cite{haug98}. Explicitly, the accumulated energy spectrum follows from
\begin{eqnarray}
\label{spectralfct_acc}
 a(\omega)=\sum_{k=0}^{nb-1}a_k(\omega)=\sum_{k=0}^{nb-1}\int_{-\infty}^{+\infty}\textup{d}\tau\, e^{i\omega t}\,a_{k}(\tau)\;,
\end{eqnarray}
where in Hartree-Fock approximation one recovers $a(\omega)=\sum_{k=0}^{n_b-1}\delta(\omega/\omega_0-\epsilon_k^0)$, compare with Fig.~\ref{Fig.intrabandsf}.

For the quasi-1D quantum dot filled with $N=3$ electrons at coupling parameters $\lambda=1$ and $\lambda=2$, Fig.~\ref{Fig.sf} shows the spectral function $a(\omega)$ including all low-energetic orbitals at different temperatures $\beta^{-1}$. The vertical solid lines indicate the discrete HF spectra $\{\epsilon_k^0\}$ and the gray filled dashed curves show $a(\omega)$ at the second Born level, being composed of the intraband functions $a_k(\omega)$ which, themselves, are represented by the thin dotted curves. Additionally, the positions of the maxima in $a_k(\omega)$ are marked by the vertical dotted lines. As a general trend, we observe a shift of almost all peaks in $a(\omega)$ compared to the HF eigenvalues $\epsilon_k^0$ of the charge carriers---in particular, the shifting is dominant around the chemical potential $\mu^0$. Moreover, the energy shifts are accompanied by a state and temperature dependent collisional broadening (finite lifetime), where, at sufficiently low temperatures, the spectral width is small and the intraband functions $a_k(\omega)$ do not overlap. Close to the ground state, less occupied HF states $k$ around $\mu^0$, typically, show larger broadening  than more strongly occupied states, see Fig.~\ref{Fig.sf}~\textbf{a}) and \textbf{d}). We note that the HOMO-LUMO gap, i.e.~the energy gap between the \underline{h}ighest \underline{o}ccupied (\underline{m}olecular or Hartree-Fock) \underline{o}rbital and the \underline{l}owest \underline{u}noccupied (\underline{m}olecular) \underline{o}rbital appearing at quasi zero temperature $\beta=10$, is reduced by electron-electron collisions, and is particularly softened at larger temperatures, compare with Fig.~\ref{Fig.sf}~\textbf{b}) and \textbf{e}) where $\beta=2$. This clearly affects the optical absorption (emission) spectra of the few-electron quantum dot\cite{brocke03}. Finally, at even higher temperatures $\beta\lesssim1$, all spectra $a(\omega)$ gradually become smooth functions with no or only few distinct peaks around the chemical potential $\mu^0$. Thereby, also low energies, essentially smaller than $\epsilon_{k=0}^0$, can be adopted by the charge carriers.

In addition, if in Eq.~(\ref{spectralfct_acc}) the intraband functions $a_k(\omega)$ are weighted by the respective occupation probabilities $n_k$ (or their inverse $\bar{n}_k=1-n_k$), one turns the energy spectrum $a(\omega)$ into $\sum_{k=0}^{n_b-1}n_k\,a_k(\omega)$ (or $\sum_{k=0}^{n_b-1}\bar{n}_k\,a_k(\omega)$). These quantities allow us to determine with which spectral weight specific electron energies are (are not) realized and thus are (are not) measurable in the correlated QD state $G^M$, see the corresponding dotted-dashed (double-dotted-dashed) curves in Fig.~\ref{Fig.sf}~\textbf{a}) to \textbf{e}).

Beyond the spectral information, the time-propagation of $G(1,2)$ also allows us to keep track of the accuracy of the correlated initial state $\vec{g}^M(\tau)$ [solution of the Dyson equation (\ref{deqif})]. Qualitatively, the accuracy can be extracted from the temporal evolution of the electron correlation energy\cite{thesis07} which is given by
\begin{eqnarray}
E_{\mathrm{corr}}(t)&=&\frac{1}{2}\int\!\textup{d}^2 r\,{I^<(\vec{r},\vec{r};t)}-E_{\mathrm{HF}}^{0}\;,\\
I^<(\vec{r}_1,\vec{r}_2;t_1)&=&-\!\int_{\cal C} \textup{d}3\,W(1-3)\,G_{12}(13;23^+)|_{t_1=t_2^+} \;,\nonumber
\end{eqnarray}
where $G_{12}$ is used in second Born approximation. When the iterative procedure discussed in Sec.~\ref{subsec:2ndB} has led to convergence and thus to a self-consistent solution $\vec{g}^M(\tau)$ of the QD Dyson equation, the correlation energy $E_{\mathrm{corr}}(t)$ must stay constant in time. Consequently, the amplitude of any small oscillatory behavior of the correlation energy (obtained by propagation) serves as a reasonable estimator for the error $\Delta E_{\mathrm{corr}}$. Throughout, with typical errors of less than $5$\%, such a test has been found to be very sensitive and useful to verify the presented results.

\section{\label{sec:conclude}Conclusion}
In this paper, we have applied the method of nonequilibrium Green's functions to study inhomogeneous strongly correlated quantum few-body systems: quantum dots with up to six spin-polarized electrons in thermodynamic equilibrium. At various interaction strengths, the self-consistent solution of the Dyson equation at the level of the second Born approximation has enabled us to focus particularly on correlation phenomena.

Close to the ground state as well as at finite temperatures, the Born approximation results yield considerable improvements for the total energies, the one-electron density, and the orbital-resolved distribution functions which give access to the electron-electron scattering processes being present in the correlated equilibrium state. Finally, the discussion of the spectral function in Sec.~\ref{subsec:spectralfct} has implied strong influence of correlations on the optical emission and absorption spectra of the considered few-electron QDs.

Of course, the second Born approximation is a very simple model. It neglects both higher order correlations (beyond second order in the interaction) and dynamical screening (e.g.~GW approximation). Nevertheless, comparison of the Born approximation results to first principle quantum Monte Carlo and exact diagonalization data suggests that this approximation is well capable to accurately describe the present system.

Further, the methods discussed in this paper should allow us to study system sizes up to $N=12$ charge-carriers in 1D and $N=6$ in 2D in a reasonable computer time on a single PC. However, the main restriction (limiting factor) is basically not the particle number itself but rather the number of basis functions, which---together with the discretized $\tau$-grid---sets up the large dimensionality of the matrices to be computed and processed. At the mean-field level, the evaluation of the two-electron integrals, Eq.~(\ref{2ei}), and the transformation of which into the HF basis are the most time-consuming parts, while solving the corresponding Dyson equation is relatively simple. For the second Born case, apart from solving the large-scale linear system (\ref{leqs}) in each iteration, in particular the computation of the self-energy $\vec{\Sigma}^\mathrm{corr}_{\lambda}(\tau)$, Eq.~(\ref{egycorr}), and the convolution integrals $\alpha_{ij}(\tau,\bar{\tau})$, Eq.~(\ref{convint}), both needed with adequate/high accuracy, account for the complexity of the calculation.

Finally and most importantly, the use of NEGFs provides a very general approach as one is capable of computing also time-dependent observables solving the KBE (Sec.~\ref{subsec:negf}) for the two-time Green's function $G(1,2)$ under nonequilibrium situations. Hence, the presented approach allows for the extension to other systems such as 'QD molecules' (assemblies of single QDs) with interdot-coupling and time-dependent carrier transport and QDs coupled to electronic leads or external (optical) laser field sources.

\begin{acknowledgments}
Part of this work was supported by the Innovationsfond Schleswig-Holstein and the Deutsche Forschungsgemeinschaft via grant FI1252/1.
\end{acknowledgments}

\appendix*
\section{Dyson equation in integral form}
In order to develop a numerical solution procedure, we, in Sec.~\ref{subsec:2ndB}, have considered the Dyson equation in its integral form which can be derived from Eq.~(\ref{deqm}) in the following way:~Multiplying Eq.~(\ref{deqm}) by the (band diagonal) Hartree-Fock state $\vec{g}^0(\bar{\bar{\tau}}-\tau)$, Eq.~(\ref{HFgf}), and integrating over $\tau\in[0,\beta]$ leads to
\begin{eqnarray}
\label{app1}
 &&\int_0^\beta\!\!\!\textup{d}\tau\,\{[-\partial_\tau-\vec{h}^0]\,\vec{g}^M\!(\tau)\}\,\vec{g}^0(\bar{\bar{\tau}}-\tau)\\
 &=&\vec{g}^0(\bar{\bar{\tau}})+\int_0^\beta\!\!\!\textup{d}\tau\!\int_0^\beta\!\!\!\textup{d}\bar{\tau}\,\{\vec{\Sigma}^M_\lambda\!(\tau-\bar{\tau})\,\vec{g}^M\!(\bar{\tau})\}\,\vec{g}^0(\bar{\bar{\tau}}-\tau)\nonumber\;.
\end{eqnarray}
Using the identity \mbox{$-\partial_\tau\{\vec{g}^M(\tau)\,\vec{g}^0(\bar{\bar{\tau}}-\tau)\}=-\partial_\tau\vec{g}^M(\tau)$} $\cdot\,\,\vec{g}^0(\bar{\bar{\tau}}-\tau)+\partial_{(\bar{\bar{\tau}}-\tau)}\vec{g}^0(\bar{\bar{\tau}}-\tau)\,\cdot\,\vec{g}^M(\tau)$, the left hand side of Eq.~(\ref{app1}) can be written as
\begin{eqnarray}
\label{app2lhs}
&&[-\vec{g}^M(\tau)\,\vec{g}^0(\bar{\bar{\tau}}-\tau)]^{\beta}_{0}\\
&+&\!\int_0^\beta\!\!\!\textup{d}\tau\,[-\partial_{(\bar{\bar{\tau}}-\tau)}\vec{g}^0(\bar{\bar{\tau}}-\tau)-\vec{h}^0\,\vec{g}^0(\bar{\bar{\tau}}-\tau)]\,\vec{g}^M\!(\tau)\nonumber\\
\label{app3lhs}
&=&\!\vec{g}^M\!(\bar{\bar{\tau}})\,+\int_0^\beta\!\!\!\textup{d}\tau\,\vec{\Sigma}_\lambda^0\,\vec{g}^0(\bar{\bar{\tau}}-\tau)\,\vec{g}^M\!(\tau)\;,
\end{eqnarray}
where the first term in Eq.~(\ref{app2lhs}) vanishes due to the anti-periodicity property of $\vec{g}^M(\tau)$ and in the second term we are allowed to insert the Dyson equation (\ref{deqHF}) for the HF reference state $\vec{g}^0(\bar{\bar{\tau}}-\tau)$. Doing so gives us expression~(\ref{app3lhs}). Finally, equating the r.h.s.~of Eq.~(\ref{app1}) with (\ref{app3lhs}) leads over to
\begin{eqnarray}
\vec{g}^M\!(\bar{\bar{\tau}})-\vec{g}^0(\bar{\bar{\tau}})&=&\int_0^\beta\!\!\!\textup{d}\tau\!\int_0^\beta\!\!\!\textup{d}\bar{\tau}\,\vec{g}^0(\bar{\bar{\tau}}-\tau)\left\{\vec{\Sigma}^M_\lambda\!(\tau-\bar{\tau})\right.\nonumber\\
&&\left.-\delta(\tau-\bar{\tau})\,\vec{\Sigma}_\lambda^0\right\}\vec{g}^M\!(\bar{\tau})\;,
\end{eqnarray}
which is equivalent to Eqs.~(\ref{deqif},\ref{mse}) in Sec.~\ref{subsec:2ndB} with the replacements ${\bar{\bar{\tau}}\leftrightarrow\tau}$, and the notations $\vec{\Sigma}_\lambda^0=\vec{\Sigma}_\lambda^s$ and $\vec{\Sigma}^r_\lambda(\tau)=\vec{\Sigma}^M_\lambda(\tau)-\delta(\tau)\,\vec{\Sigma}^s_\lambda\;,$




\begin{thebibliography}{100}

\bibitem{jacak98} L. Jacak, P. Hawrylak, and A. W\'{o}js, \textit{Quantum Dots}, (Springer, Berlin Heidelberg, 1998).

\bibitem{ashoori96} R.C. Ashoori, Nature (London) \textbf{379}, 413 (1996).

\bibitem{brocke03} T. Brocke, M.-T. Bootsmann, M. Tews, B. Wunsch, D. Pfannkuche, Ch. Heyn, W. Hansen, D. Heitmann, and C. Sch\"uller, Phys. Rev. Lett. \textbf{91}, 25 (2003).

\bibitem{baer04} N. Baer, P. Gartner, and F. Jahnke, Eur. Phys. J. B \textbf{42}, 231 (2004).

\bibitem{szafran04} B. Szafran, F.M. Peeters, S. Bednarek, and J. Adamowski, Phys. Rev. B \textbf{69}, 125344 (2004).

\bibitem{indlekofer05} K.M. Indlekofer, J. Knoch, and J. Appenzeller , Phys. Rev. B \textbf{72}, 125308 (2005).

\bibitem{banyai93} L. Banyai, and S. W. Koch, \textit{Semiconductor Quantum Dots} (World Scientific, Singapore, 1993).

\bibitem{reimann02} S.M. Reimann, and M. Manninen, Rev. Mod. Phys. \textbf{74}, 1283 (2002).

\bibitem{kvaal07} S. Kvaal, M. Hjorth-Jensen, and H.M. Nilsen, Phys. Rev. B \textbf{76}, 085421 (2007).

\bibitem{ciftja06} O. Ciftjy, and M.G. Faruk, J. Phys.: Condens. Matter \textbf{12}, 2623-2633 (2006).

\bibitem{jauregui93} K. Jauregui, W. H\"ausler, and B. Kramer, Europhys. Lett. \textbf{24}, 581-587 (1993).

\bibitem{yannou07} C. Yannouleas, and U. Landman, Rep. Prog. Phys. \textbf{70}, 2067-2148 (2007).

\bibitem{ludwig08} P. Ludwig, K. Balzer, A. Filinov, H. Stolz, and M. Bonitz, New. J. Phys. \textbf{10}, 083031 (2008).

\bibitem{reusch01} B. Reusch, W. H\"ausler, and H. Grabert, Phys. Rev. B \textbf{63}, 113313 (2001); B. Reusch and H. Grabert, Phys. Rev. B \textbf{68}, 045309 (2003).

\bibitem{rontani06} M. Rontani, C. Cavazzoni, D. Bellucci, and G. Goldoni, J. Chem. Phys. \textbf{124}, 124102 (2006).

\bibitem{egger99} R. Egger, W. H\"ausler, C.H. Mak, and H. Grabert,  Phys. Rev. Lett. \textbf{82}, 16 (1999).

\bibitem{filinov01} A.V. Filinov, M. Bonitz, and Yu. E. Lozovik, Phys. Rev. Lett. \textbf{86}, 17 (2001).

\bibitem{filinov04} A.V. Filinov, C. Riva, F.M. Peeters, Yu.E. Lozovik, and M. Bonitz, Phys. Rev. B \textbf{70}, 035323 (2004).

\bibitem{bracker05} A.S. Bracker, E.A. Stinaff, D. Gammon, M.E. Ware, J.G. Tischler, D. Park, D. Gershoni, A.V. Filinov, M. Bonitz, F.M.Peeters, and C. Riva, Phys. Rev. B \textbf{72}, 035332 (2005) 

\bibitem{dahlen05} N.E. Dahlen, and R. van Leeuwen, J. Chem. Phys. \textbf{122}, 164102 (2005).

\bibitem{dahlen06} N.E. Dahlen, R. van Leeuwen, and A. Stan, J. Phys: Conf. Ser. \textbf{35}, 340-348 (2006).

\bibitem{dahlen07} N.E. Dahlen, and R. van Leeuwen, Phys. Rev. Lett. \textbf{98}, 153004 (2007).

\bibitem{bonitz96} M. Bonitz, D. Kremp, D.C. Scott, R. Binder, W.D. Kraeft, and H.S. K\"ohler, J. Phys.: Condens. Matter \textbf{8}, 6057 (1996).

\bibitem{binder97} R. Binder, H.S. K\"ohler, and M. Bonitz, Phys. Rev. B \textbf{55}, 5110 (1997).

\bibitem{kwong98}N.H. Kwong, M. Bonitz, R. Binder and S. K\"ohler, phys. stat. sol. (b) \textbf{206}, 197 (1998).

\bibitem{kadanoff62} L.P. Kadanoff, and G. Baym, \textit{Quantum Statistical Mechanics} (Benjamin, Inc., New York, 1962).

\bibitem{martin59} P.C. Martin, and J. Schwinger, Phys. Rev. \textbf{115}, 1342 (1959).

\bibitem{baym62} G. Baym, Phys. Rev. \textbf{127}, 1391 (1962).

\bibitem{kubo5766} R. Kubo, J. Phys. Soc. Jpn. \textbf{12}, 570 (1957); Rep. Prog. Phys. \textbf{29}, 255 (1966).

\bibitem{ku02} W. Ku, and A.G. Eguiluz, Phys. Rev. Lett. \textbf{89}, 126401 (2002).

\bibitem{echenique07} P. Echenique, J.L. Alonso, Mol. Phys. \textbf{105}, 3057-3098 (2007).

\bibitem{roothaan51} C.C.J. Roothaan, Rev. Mod. Phys. \textbf{20}, 69 (1951); G.G. Hall, Proc. R. Soc. (London) \textbf{A205}, 451 (1951).

\bibitem{upmesh} The uniform power mesh discretizes the half-interval $[-\beta,-\frac{\beta}{2}]$ ($[-\frac{\beta}{2},0]$) by $p-1$ bisections of the most lower (upper) part and introduces, in each division, an evenly spaced grid of $u$ subdivisions. With the integers $u$ and $p$ the total number of grid points is $n_m=2up+1$ and the smallest (largest) mesh-spacing is $\frac{1}{2}\beta/(2^{p-1} u)$ ($\beta/(4u)$).

\bibitem{thesis07} K. Balzer, Diploma thesis, Kiel University (2007).

\bibitem{mikhailov02} S.A. Mikhailov, Phys. Rev. B \textbf{65}, 115312 (2002).

\bibitem{QMC}{Quantum Monte Carlo [e.g.~path integral Monte Carlo simulations\cite{egger99,filinov01,filinov04}], as a \textit{first-principle} many-body approach, accounts for all correlation effects and yields numerically exact results.}

\bibitem{tran01} M.N. Tran, M.V.N. Murthy, and R.K. Bhaduri, Phys. Rev. E \textbf{63}, 031105 (2001).

\bibitem{merkt91} U. Merkt, J. Huser, and M. Wagner, Phys. Rev. B \textbf{43}, 7320 (1991).

\bibitem{giovannini07} U. De Giovannini, F. Cavaliere, R. Cenni, M. Sassetti, and B. Kramer, New. J. Phys. \textbf{9}, 93 (2007).

\bibitem{bonitz99} M. Bonitz, D. Semkat, and H. Haug, Eur. Phys. J. B \textbf{9}, 309-314 (1999).

\bibitem{haug98} H. Haug, L. Banyai, Solid State Comm. \textbf{100}, 303 (1998).

\end{thebibliography}
\end{document}